\documentclass[superscriptaddress,twocolumn,pre]{revtex4}

\usepackage{ifthen}
\newboolean{pnas}
\setboolean{pnas}{false}
\newcommand{\siref}{SI Text}

\usepackage{amsmath}
\usepackage{amsfonts}
\usepackage{amssymb}
\usepackage{mathtools}
\usepackage{graphicx}
\graphicspath{{images/}}
\usepackage{color}
\usepackage[pdfstartview=FitH,
            breaklinks=true,
            bookmarksopen=false,
            bookmarksnumbered=true,
            colorlinks=true,
            linkcolor=black,
            citecolor=black,
            urlcolor=black,
            pdftitle={},
            pdfauthor={Andreas Mayer},
            pdfsubject={}
            ]{hyperref}

\newcommand{\B}{\boldsymbol}
\newcommand{\ud}{\mathrm{d}}
\newcommand{\beq}{\begin{equation}}
\newcommand{\eeq}{\end{equation}}

\newcommand{\<}{\langle}
\renewcommand{\>}{\rangle}

\def\(({\left(}
\def\)){\right)}                       
\def\[[{\left[}
\def\]]{\right]}

\DeclareMathOperator*{\argmin}{\arg\!\min}
\DeclareMathOperator\artanh{artanh}

\newcommand{\AM}[1]{{\color{black}#1}}
\newcommand{\AW}[1]{{\color{black}#1}}

\usepackage{adjustbox}
\adjustboxset{max width=1.15\textwidth}

\begin{document}

\title{How a well-adapting immune system remembers}
\author{Andreas Mayer}
\affiliation{Lewis-Sigler Institute for Integrative Genomics,
Princeton University, Princeton, NJ 08544, USA}
\affiliation{Laboratoire de physique th\'eorique,
    CNRS, Sorbonne Universit\'e and \'Ecole Normale Sup\'erieure (PSL),
    75005 Paris, France}
\author{Vijay Balasubramanian}
\affiliation{David Rittenhouse Laboratories, University of Pennsylvania, Philadelphia,
    PA 19104;\\
    and Initiative for the Theoretical Sciences, The
    Graduate Center, The City University of New York, NY 10016, USA}
\author{Aleksandra M. Walczak}
\thanks{These authors contributed equally. Please send
  correspondence to \url{tmora@lps.ens.fr}, \url{awalczak@lpt.ens.fr}}
\affiliation{Laboratoire de physique th\'eorique,
    CNRS, Sorbonne Universit\'e and \'Ecole Normale Sup\'erieure (PSL),
    75005 Paris, France}
\author{Thierry Mora}
\thanks{These authors contributed equally. Please send
  correspondence to \url{tmora@lps.ens.fr}, \url{awalczak@lpt.ens.fr}}
\affiliation{Laboratoire de physique statistique,
    CNRS, Sorbonne Universit\'e, Universit\'e Paris-Diderot, and \'Ecole normale sup\'erieure (PSL),
    75005 Paris, France}

\begin{abstract}
     % abstract
An adaptive agent predicting the future state of an environment must weigh trust in new observations against prior experiences. In this light, we propose a view of the adaptive immune system as a dynamic Bayesian machinery that updates its memory repertoire by balancing evidence from new pathogen encounters against past experience of infection to predict and prepare for future threats. This framework links the observed initial rapid increase of the memory pool early in life followed by a mid-life plateau to the ease of learning salient features of sparse environments. We also derive a modulated memory pool update rule in agreement with current vaccine response experiments. Our results suggest that pathogenic environments are sparse and that memory repertoires significantly decrease infection costs even with moderate sampling. The predicted optimal update scheme maps onto commonly considered competitive dynamics for antigen receptors.

\end{abstract}

\maketitle

 % intro
\ifthenelse{\boolean{pnas}}{}{\section{Introduction}}
\ifthenelse{\boolean{pnas}}{\dropcap{A}}{A}ll living systems sense the environment, learn from the past, and adapt predictively to prepare for the future.  Their task is challenging because environments change constantly, and it is impossible to sample them completely.  Thus a key question is how much weight should be given to new observations versus accumulated past experience. Because evidence from the world is generally uncertain, it is convenient to cast this problem in the language of probabilistic inference where past experience is encapsulated in a prior probability distribution which is updated according to sampled evidence.  This framework has been successfully used to understand aspects of cellular~\cite{Perkins2009,Kobayashi2010,Siggia2013,Sivak2014}  and neural~\cite{DeWeese1998,Beck2008,Deneve2008a,Wark2009} sensing.  Here, we propose that the dynamics of the adaptive immune repertoires of vertebrates can be similarly understood as a system for \AM{probabilistic inference of pathogen statistics}.

The adaptive immune system relies on a diverse repertoire of B and T cell receptors to protect the host organism from a wide range of pathogens. These receptors are expressed on clones of receptor-carrying cells present in varying copy numbers. \AM{A defining feature of the adaptive immune system is its ability to change its clone composition throughout the lifetime of an individual, in particular via the formation of memory repertoires of B and T cells following pathogen encounters \cite{Janeway,Farmer1986,Ahmed1996,Perelson1997,Farber2014,Buchholz2016}.
In detail, after a proliferation event that follows successful recognition of a foreign antigen, some cells of the newly expanded clone acquire a memory phenotype. These cells make up the memory repertoire compartment that is governed by its own homeostasis, separate from the inexperienced naive cells from which they came. Upon reinfection by a similar antigen, memory guarantees a fast immune response. With time, our immune repertoire thus becomes specific to the history of infections, and adapted to the environments we live in.}
However, the commitment of part of the repertoire to maintaining memory must be balanced against the need to also provide broad protection from as yet unseen threats. What is more, memory will lose its usefulness over time as pathogens evolve to evade recognition.

How much benefit can immunological memory provide to an organism?   How much memory should be kept to minimize harm from infections?  How much should each pathogen encounter affect the distribution of receptor clones?  To answer these questions we extend a framework for predicting optimal repertoires given pathogen statistics \cite{Mayer2015} by explicitly considering the inference of pathogen frequencies as a Bayesian forecasting problem~\cite{Chen2003}. \AM{We derive the optimal repertoire dynamics in a temporally varying environment. This approach can complement more mechanistic studies of the dynamics and regulation of immune responses \cite{Perelson1997,Deem2003,Antia2005,Chakraborty2017} by revealing adaptive rationales underlying particular features of the dynamics.
In particular, we link the amount of memory production to the variability of the environment and show that there exists an optimal timescale for memory attrition. Additionally, we demonstrate how biologically realistic population dynamics can approximate the optimal inference process, and analyze conditions under which memory provides a benefit.  Comparing predictions of our theory to experiment, we argue for a view in which the adaptive immune system can be interpreted as a machinery for learning a highly sparse distribution of antigens.}

 % setup
\ifthenelse{\boolean{pnas}}{\section*{Theory of optimal immune prediction}}{\section{Theory of optimal immune prediction}}

\begin{figure*}
    \begin{adjustbox}{center}
    \includegraphics[scale=1.2]{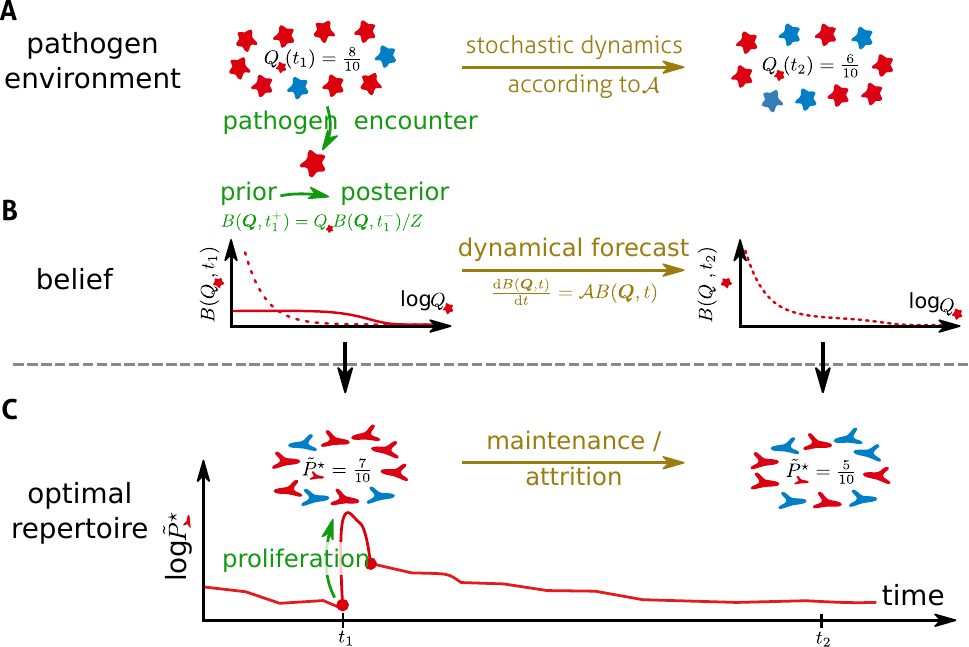}
    \end{adjustbox}
    \caption{Sketch of a model of immune repertoire dynamics as a sequential inference process about a time-varying pathogen distribution.
    (A) The organism lives in a pathogenic environment with frequencies of different pathogen strains that change over time.
    (B) Past pathogen encounters provide an avenue for the immune system to learn the pathogen distribution. Using sequential Bayesian inference provides an optimal way to update the beliefs about the frequencies of different pathogens over time. 
    (C) Bases on its beliefs about the prevalence of pathogens the optimal immune dynamics allocates lymphocytes across different pathogens to minimizes the expected harm from infections. Broadly, the more frequent a pathogen is the more the organism should be covered. This resource allocation maps the changes in beliefs to the changes in the repertoire composition.
    \label{figsetup}
    }
\end{figure*}

The pathogenic environment is enormous and the immune system can only sample it sparsely, as  pathogens enter into contact with it at some rate $\lambda$.
We consider an antigenic space of $K$ different pathogens with time-varying frequencies $\B Q(t) = (Q_1(t), ..., Q_K(t))$. These frequencies are unknown to the organism, and evolve stochastically. Their dynamics is formally described by a \AM{Fokker-Planck operator $\mathcal A$} encoding how pathogenic frequencies change (Fig.~\ref{figsetup}A and Methods).
We reason that the immune system should efficiently use the information available through these encounters, along with  prior knowledge of how pathogens evolve encoded in the system dynamics, to build an internal representation of the environment (Fig.~\ref{figsetup}B).
\AM{Biologically, we can think about this representation as being encoded in the composition of the adaptive immune repertoire (the size and specificity of naive and memory lymphocyte clones), but generally further cellular memory mechanisms might also contribute.} Based on this representation of the world, the immune system should organize its defenses to minimize harm from future infections (Fig.~\ref{figsetup}C).

\AM{How could the immune system leverage a representation of beliefs about pathogen frequencies to provide effective immunity?
Each lymphocyte (B or T cell) of the adaptive immune system expresses on its surface a single receptor $r$ out of $L$ possible receptors.  This receptor endows the lymphocyte with the ability to specifically recognize pathogens  (labeled $a$) with probability $f_{a,r}$.
The immune repertoire is defined by frequencies of these receptors across the lymphocyte population, denoted by $\B P = (P_1, ... P_L)$. These frequencies sum up to one, which implies a resource allocation trade-off between the different receptor types -- having more of one in the repertoire implies having less of others.
How much harm an infection inflicts depends on how much resources the immune system has devoted to fighting the infection, i.e. the fraction $\tilde P_a(t) = \sum_r f_{a,r} \, P_r$ of the repertoire specific to antigen $a$, which we will refer to as the {\it coverage} of the antigen. Given the pathogen frequencies $\B Q(t)$ and repertoire distribution $\B P(t)$, the mean harm cause by the next infection is given by $\sum_a Q_a\cdot c(\tilde P_a)$, where $c$ is decreasing function of the fraction of the repertoire specific to the infection \cite{Mayer2015}. The host organism does not know $\B Q$ with certainty, but has an internal belief $B(\B Q,t)$ about the frequencies learned through sampling during previous infections.
An optimal immune system can then distribute its resources to minimize the expected harm of future of infections:
\begin{equation} \label{eqoptP}
    \B P^\star(t) = \argmin_{\B P} \sum_a \hat{Q}_a(t)\cdot c(\tilde P_a) \equiv G(\hat{\B Q}(t)),
\end{equation}
where $\hat {\B Q}(t)\equiv \<\B Q\>_{B(\B Q, t)}$ are the expected frequencies of pathogens.
Although the function $G$ may be complicated, it generally implies that receptors that are specific to frequent infections (high $\hat Q_a$) should be well represented in the optimal repertoire (high $\tilde P_a^\star$) (\cite{Mayer2015} and Methods).
In this framework we have assumed that infections, their clearing by the immune system, and the subsequent update of the repertoire, are all fast compared to changes in the environment, which occur  over a time $\tau$, and to the mean time between pathogenic encounters ($\lambda^{-1}$).
}

The internal representation of the environment can be regarded as a system of beliefs, or guesses, about pathogen frequencies.
Formally, these beliefs can be represented in the form of a probability distribution function $B(\B Q,t)$ over pathogen frequencies, which the host implicitly computes using all the information it has garnered over time. 
Optimally, these beliefs are computed by the rules of Bayesian sequential forecasting, by combining the memory of past encounters with knowledge of the stochastic rules under which the pathogenic environment evolves (Methods).
Optimally, the belief distribution should be initialized at birth to reflect 
the steady state distribution of the dynamics,
\begin{equation} \label{eqprior}
    B(\B Q,0) = \rho_s(\B Q),
\end{equation}
where $\mathcal{A}\rho_s=0$.
Upon encountering a pathogen $a$ at time $t$, \AM{the prior belief distribution $B(\B Q,t^-)$ is combined with the likelihood of the observed pathogen $Q_a$ to compute the post-encounter belief $B(\B Q,t^+)$ according to Bayes rule \cite{Chen2003}:}
\begin{equation} \label{equpdate}
    B(\B Q, t^+) = \frac{Q_a \, B(\B Q, t^-)}{\int \ud \B Q' \;  Q'_a \, B(\B Q', t^-)}.
\end{equation}
Between encounters, \AM{the immune system should continue to update its beliefs by forecasting how pathogen frequencies change with time. The optimal way to do so is to project the old belief distribution forward in time using \cite{Chen2003}}
\beq \label{eqpred}
\frac{\ud B(\B Q, t)}{\ud t}=\mathcal{A}B(\B Q, t).
\eeq
This prediction step, which is performed in the absence of any new information, relies on the immune system implicitly ``knowing'' the probability laws governing the stochastic evolution of the environment---but not, of course, the actual path that it takes.
\AM{In the results section we show how Eqs.~\ref{eqprior}-\ref{eqpred} can be turned from abstract belief updates into dynamical equations for a well-adapting immune repertoire.}

\AM{
The Bayesian forecasting framework provides a broad account of the possible adaptive value of many features of the adaptive immune system \AM{without the need for additional assumptions}.
Immune memory formed after a pathogenic challenge is explained as an increase in optimal protection level resulting from an increase in estimated pathogen frequency, following Eq.~\ref{equpdate}.
Attrition of immune memory is also adaptive, because it allows the immune repertoire to forget about previously seen pathogens which it should do in a dynamically changing environment (Eq.~\ref{eqpred}).
Lastly, some of the biases in the recombination machinery and initial selection mechanisms \cite{Mora2016} represent an evolutionary prior (Eq.~\ref{eqprior}) which tilts the naive repertoire towards important regions of antigenic space.

We are proposing an interpretive framework for understanding adaptive immunity as a scheme of sequential inference.
This view provides two key insights. First, it \AW{confirms the intuition that} new experience should be balanced against  previous memory and against unknown threats in order for adaptive immunity to work well. Second, it suggests a particular dynamics of implicit belief updates that can globally reorganize the immune repertoire to minimize harm from the pathogenic environment. 
Going beyond these broad ideas, in the Results section we analyze in details a model for optimal immune prediction in which all these statements can be made mathematically precise. We also show a plausible \AW{implementation} that the immune system could follow to approximate optimal Bayesian inference, and we compare the resulting dynamics with specific features of the adaptive immune system.

}

 % results
\ifthenelse{\boolean{pnas}}{\section*{Results}}{\section{Results}}

\ifthenelse{\boolean{pnas}}{\subsection*{A lymphocyte dynamics for approximating optimal sequential inference}}{\subsection{A lymphocyte dynamics for approximating optimal sequential inference}}

For concreteness we consider a \AM{drift-diffusion model of environmental change (Eq.~\ref{eqwfdiffusion} in Methods). The drift-diffusion model, while clearly a much simplified model of real evolution, captures two key features of changing pathogenic environments: the co-existence of diverse pathogens, and the temporal turnover of dominant pathogen strains. The drift-diffusion model is mathematically equivalent to a classical neutral stochastic evolution of pathogens \cite{Kimura1964} driven by genetic drift happening on a characteristic timescale $\tau$ and immigration from an external pool with  immigration parameters $\B \theta=(\theta_1,\ldots,\theta_K)$ (Eq.~\ref{eqwfdiffusion} in Methods).
Generally, pathogens are under selective pressure to evade host immunity, and strains are replaced faster than under the sole action of genetic drift. Matching the timescale of pathogen change to those observed experimentally, the model then provides a simple, effective description of the pathogen dynamics.
}

In this case, we show that optimal Bayesian belief update dynamics can be approximated by maintaining a memory of an effective count of previous encounters  $\B n=(n_1,\ldots,n_K)$, initialized to the immigration rates $\B n(0)=\B\theta$, and
subject to the update rules (see \siref~\ref{appprobaexpansion}):
\begin{align}
    n_a(t^+) &= n_a(t^-) + 1\quad\text{upon encountering }a \label{equpdaten} \\
    \tau\frac{\ud \B n}{\ud t} &= - \frac{1}{2} \(( |\B n| - 1\)) \((\B n - \B \theta\)), \label{eqpredn}
\end{align}
where $|\B n|=\sum_{a'} n_{a'}$. The expected frequency of each pathogen (used in Eq.~\ref{eqoptP}) is estimated from these counts as:
\begin{equation} \label{eqavQ}
    \hat Q_a(t)\approx  \frac{n_a}{|\B n|},
\end{equation}
We checked the accuracy of this approximation explicitly by comparing it to an exact solution computed by spectrally expanding the generator of the stochastic dynamics (see \siref~\ref{appspectralexpansion} and Fig.~\ref{figdyncomp}).

An optimal immune system should should map the counts above into a receptor repertoire ${\B P}^\star$ as in Eq.~\ref{eqoptP}. \AM{The repertoire then follows a dynamics derived from Eqs.~\ref{equpdaten}-\ref{eqpredn} (see \siref~\ref{appinducedrepdyn}), in which memory of past infections is encoded only in the repertoire composition itself and in a single global variable representing the total memory that is kept (encoding $|\B n|$).}
To understand this, consider  a cost function $c(\tilde P_a) = - \log \tilde P_a$ and uniquely specific receptors, $f_{a,r} = \delta_{a,r}$.  In this case  the mapping Eq.~\ref{eqoptP} is the identity, i.e. ${\B P^\star}(t)  = \hat{\B Q}(t)$ (\cite{Mayer2015} and Methods).  Then the optimal repertoire dynamics can be achieved simply by having clone sizes of different receptors follow Eq.~\ref{equpdaten},\ref{eqpredn} up to some scaling. More generally the optimal repertoire is some non-linear mapping of the encounter counts, but only requires information that can be represented in population sizes of different clones, which are quantities  regulated by the actual biological dynamics.

\ifthenelse{\boolean{pnas}}{\subsection*{Learnability of pathogen distribution implies a sparse pathogenic landscape}}{\subsection{Learnability of pathogen distribution implies a sparse pathogenic landscape}}
\label{reslearnability}

The immune system must be prepared to protect us not just from one pathogen but a whole distribution of them. Even restricting recognition to short peptides and accounting for cross-reactivity \cite{Mason1998}, estimates based on precursor frequencies for common viruses give an effective antigen environment of size $K\sim 10^5$--$10^7$ \cite{Zarnitsyna2013}. How can the immune system learn anything useful about such a high dimensional distribution from a limited number of pathogenic encounters? Naively, one might expect that the number of samples needed to learn the distribution of pathogens must be larger than of the number of pathogens, i.e., $\lambda t\sim K$, where $t$ is the time over which learning takes place. Although little is known about the receptor-antigen encounter rate $\lambda$, this estimate suggests that the pathogenic environment is not easily learnable and therefore memory has limited utility

This apparent paradox can be resolved by the fact that the pathogenic environment may be sparse, meaning that only a small fraction of the possible pathogens are present at any given time. In \AM{our model of the pathogen dynamics}, this sparsity is controlled by the parameter $\B \theta$.  In the scenario that we are considering, typical pathogen landscapes $\B Q$ are drawn from the steady state distribution $\rho_s(\B Q)$ of the immigration-drift dynamics, which is a Dirichlet distribution parametrized by $\B\theta$ (Eq.\ref{eqdirichlet} in \siref~\ref{appwfdiffusion}).
When $\theta_a$ is small, the distribution is peaked at $Q_a=0$, meaning that pathogen $a$ is absent a majority of the time. For instance, for uniform $\theta_a\equiv\theta\ll 1$, the effective number of pathogens present at any given time is $K\theta$
(see \siref~\ref{appscalingextreme}). Since the system only needs to learn about the pathogens that are present, the condition for efficient learning should naively be $\lambda t \sim K\theta$, which is much easier to achieve realistically for small $\theta$.

\begin{figure*}
    \begin{adjustbox}{center}
    \includegraphics{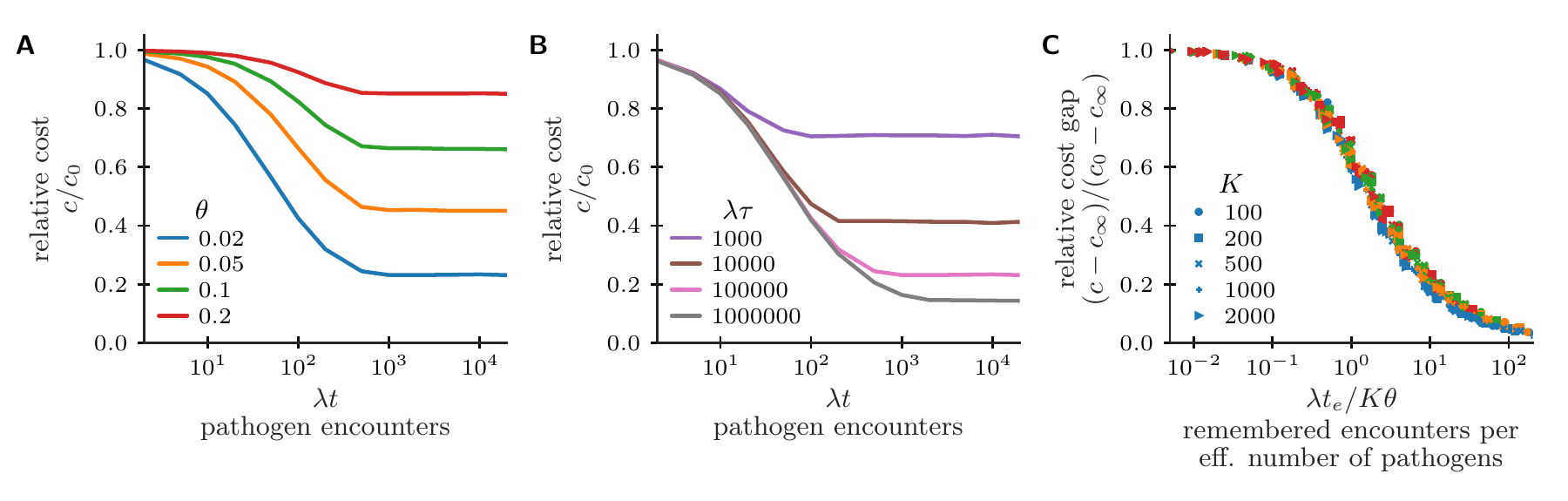}
    \end{adjustbox}
    \caption{Advantage of immunological memory depends on sufficient sampling. The mean expected cost of an infection in a changing environment is a function of both the age of the organism $t$, the timescale $\tau$ on which the environment changes, and the sparsity $1/\theta$ of the environment. (A) Relative cost as a function of age for environments with different sparsity (for fixed $\lambda \tau = 10^5$, $K = 2000$). (B) Relative cost as a function of age for environments changing more or less rapidly (for fixed $\theta=0.02$, $K = 2000$). (C) Collapse of the data by plotting the relative cost gap against the number of samples per effective dimension.
    Simulations with logarithmically spaced parameters in the ranges $\lambda t$ from 1 to 5000, $\lambda \tau$ from 1 to 1000000, $K$ from 100 to 2000, $\theta$ from 0.01 to 0.2.
    Cost is $c(\tilde P_a) = 1/\tilde P_a$ and receptors are assumed to be uniquely specific, $f_{a,r} = \delta_{a,r}$. 
    \label{figdynamicscaling}
    }
\end{figure*}

Our theory can be used to quantify the benefit of memory as a function of the different immunological parameters.
We compute the optimized cost function $c(t)=\sum_a Q_a(t)c(\tilde P_a^\star(t))$, and study how it decreases as a function of age, $t$, relative to the cost at birth $c_0=c(t=0)$, as the organism learns from pathogen encounters. This relative cost also depends on the encounter rate $\lambda$, the size of the pathogenic space $K$, the sparsity of pathogenic space $\theta$ and timescale of change in the environment $\tau$. 
Fig.~\ref{figdynamicscaling}A shows that the benefit of memory increases with pathogen sparsity -- when $\theta$ is small,  even a few encounters suffice to seed enough memory to reduce the cost of future infections.  The cost saturates with age to a value $c_\infty$, either because memory approaches optimality, or because memory eventually gets discarded and renewed as the environment changes. Fast changing environments lead to an earlier and higher saturation of the cost with age (Fig.\ref{figdynamicscaling}B) since learning and prediction are limited by decorrelation of the environment.
\AM{The pathogen dynamics is sped up when there are strong selection pressures to evade immunity. Faster dynamics decrease learning efficiency and in turn reduce selective pressures. The effective timescale should in practice be set by a co-evolutionary balance between both effects.}

Analytical arguments show that in the limit of few samples the relative cost $c/c_0$ achievable in a static environment scales as $\lambda t/K \theta$ (\siref~\ref{appscaling}). In general, we find that the cost is a function of $\lambda t_e/K\theta$, where the effective time $t_e$ is defined via $ \lambda t_e = |\B n(t)| - |\B n(0)| \approx \sqrt{2\lambda \tau}\tanh(\lambda t/\sqrt{2\lambda\tau})$ with $\B n(t)$ being the vector of the encounter counts discussed above  (see \siref~\ref{appdynamicscaling} for derivation from Eqs.~\ref{equpdaten}-\ref{eqpredn}). Plotted in terms of this variable the relative cost gap as function of time, $(c - c_\infty)/ (c_0 - c_\infty)$ collapses onto a single curve for all parameter choices (Fig.~\ref{figdynamicscaling}C).  Fig.~\ref{figdynamicscaling}C shows that the cost drops by a factor of $\sim 2$, when $ \lambda t_e/K\theta \sim 1$.  Thus, there is a substantial benefit to memory already when the effective number of encounters is comparable to the effective number of pathogens.  At young ages (small $t$) or with slowly changing environments (large $\tau$),  $t_e \approx t$ and so this condition is simply $\lambda t \sim K\theta$, i.e. the total number of encounters should be comparable to effective number of pathogens that are present. 

\AM{

\ifthenelse{\boolean{pnas}}{\subsection*{Optimal attrition timescale}}{\subsection{Optimal attrition timescale}}

Our theory suggests that there is an optimal timescale for forgetting about old infections which is \AM{related to the timescale} over which the environment varies.
Eq.~\ref{eqpredn} shows that memory should optimally be discounted on an effective timescale $\tau_{\rm mem} = 2 \tau/(|\B n| -1)$. Comparing this to the slowest timescale of environmental variation, $\tau_c = 2\tau/|\B \theta|$ (Eq.~\ref{deftauc}), where $|\B \theta| = \sum_a \theta_a$, we have
\begin{equation} \label{eqtaumem}
    \tau_{\rm mem} = \tau_c \frac{|\B \theta|}{|\B n| - 1}.
\end{equation}
The timescale on which old memories should be forgotten scales with the environmental correlation timescale. The two timescales are equivalent when the immune system has little information about the pathogenic environment ($|\B n| \sim |\B \theta|$).
Given the long timescales over which many relevant pathogens change, immune memory should generally be long-lived (with the timescale of decay being of the order of years or decades). Indeed, despite the relatively short life span of memory cells \cite{Antia1998}, constant balanced turnover keeps elevated levels of protection for decades after an infection, even  in the absence of persistent antigens \cite{Amanna2007,Sallusto2010,Macallan2017}.

Interestingly, our theory predicts that memory should be discounted more quickly when the immune system has gathered more information (larger $|\B n|$).
\AM{Using the mean-field equations for $|\B n(t)| - |\B n(0)|$ from Results~\ref{reslearnability} we can derive how the memory time scales at steady state at high sampling rate. Using that for large times $|\B n(t)| \gg |\B n(0)|$ holds in the high sampling rate limit, one can simplify the mean-field result to $|\B n| \sim \sqrt{2 \lambda \tau}$ in steady state ($t\to \infty$). Combined with Eq.~\ref{eqtaumem} $\tau_{\rm mem} = \sqrt{\tau_c |\B \theta| / \lambda}$ follows, which shows that a larger sampling rate leads to a faster discounting of past evidence. } This is reminiscent of results in optimal cellular signalling where there are similar trade-offs between noise averaging and responsiveness to changes in the input signal \cite{Becker2015}.
}
 
\ifthenelse{\boolean{pnas}}{\subsection*{Memory production in sparse environments should be large and decrease with prior exposure}}{\subsection{Memory production in sparse environments should be large and decrease with prior exposure}}

The theory can be used to make quantitative and testable predictions about the change in the level of protection that should follow a pathogen encounter. Consider an infection cost function that depends as a power law on the coverage, $c(\tilde P_a) = 1/\tilde P_a^\alpha$,  with a {\it cost exponent} $\alpha$ that sets how much attention the immune system should pay to recognizing rare threats.    (Below we will use the shorthand $\alpha = 0$ to indicate logarithmic cost.) 
Cost functions of this form can be motivated by considering the time to recognition of an exponentially growing antigen population by the immune system \cite{Mayer2015}, or, alternatively, by considering the time delay of the expansion of the precursor cells to some fixed number of effector cells (\siref~\ref{applogcost}).

In the simplest model for repertoire updates recognition of pathogens leads to proliferation proportionally to the number of specific precursor cells, followed by a homeostatic decrease of the memory pool \cite{Antia2005, DeBoer2013}. Thus the fold-change ${\tilde P_a(t^+)}/{\tilde P_a(t^-)}=\textrm{const}$ where $t^-$, $t^+$ are times just before and after the encounter. By contrast, our Bayesian theory predicts that the fold change upon encountering pathogen $a$ should be
\begin{equation}\label{eqfoldchange}
    \frac{\tilde P_a^\star(t^+)}{\tilde P_a^\star(t^-)} = \((1 + {\kappa}/{\tilde P_a^\star(t^-)^{(1+\alpha)}}\))^{1/(1+\alpha)},
\end{equation}
where  $\kappa$ depends on prior expectations about the antigenic environment and previous pathogen encounters (see Methods~\ref{methodsupdating}). Setting $\alpha = 0$ gives the result for a logarithmic cost function.

\begin{figure}
    \centering
    \includegraphics[scale=0.9]{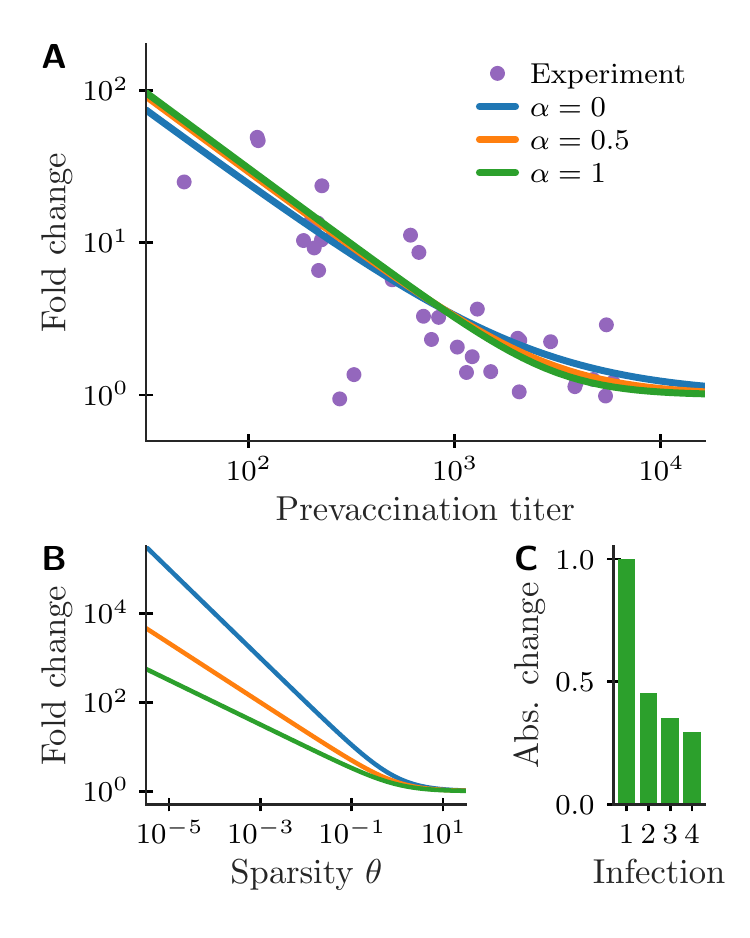}
    \caption{
        Changes in protection levels upon infections for cost functions $c(\tilde P_a) = 1/\tilde P_a^\alpha$ ($\alpha = 0$ indicates logarithmic cost).
        (A) The prediction of the Bayesian model (Eq.~\ref{eqfoldchange} for different $\alpha$) closely fits experimental data on antibody titers, while a constant fold change model (which would correspond to a straight horizontal line) does not.
        Data on pre- and post-vaccination antibody titers against stem and head hemaglutinin epitopes following a booster vaccination with inactivated H5N1 in humans from \cite{Ellebedy2014}. 
        (B) Optimal fold change of coverage for $n_i(t^-) = \theta$ for all $i$. The fold change increases with the sparsity of the environment controlled by parameter $\theta$.
        (C) Absolute change in coverage for primary infection and reinfections (for $\alpha=1$) normalized such that the change for the primary infection is 1 and neglecting attrition.
    \label{figboost}
    }
\end{figure}

To understand this prediction first consider the effect of a primary infection on a naive repertoire, $\theta_a\equiv \theta$, $\tilde P^\star(0)=1/K$, and $|\B n(0)|=K\theta$ where the receptors are uniquely specific ($f_{a,r} = \delta_{a,r}$). In this case $\kappa = 1/K^{1+\alpha}\theta$ (see \siref~\ref{appupdating}) and Eq.~\ref{eqfoldchange} predicts a fold-change of $(1+1/\theta)^{1/(1+\alpha)}$. \AM{We have argued previously that their learnabilility implies that pathogenic environments are sparse, i.e. $\theta \ll 1$. Therefore we predict that primary antigenic encounter should lead to a large memory production.  Experimentally, memory production typically leads to the proliferation of antigen-specific cells by a factor of 100-1000-fold \cite{Buchholz2016}, in qualitative agreement with this prediction. Turning the argument around, such a large increase in protection upon an encounter is only adaptive in highly sparse environments. Quantitatively, it implies a sparsity parameter $\theta \sim 10^{-6}$--$10^{-4}$ (here taking $\alpha=1$ for definiteness) (Fig.~\ref{figboost}B).} Combined with the estimate $K\sim 10^5$--$10^7$ \cite{Zarnitsyna2013} this suggests that the effective number of pathogens at any given time ranges from $K\theta=0.1$ to $1,000$.

To test Eq.~\ref{eqfoldchange} on immunological data, we fit the Bayesian update model to experiments reporting fold-changes in antigen titers upon booster vaccinations against influenza from \cite{Ellebedy2014} (Fig.~\ref{figboost}A) using least-squares. Titers correspond to the concentration of antibodies that are specific to the antigen $a$, and can thus be viewed as an experimental estimate of $\tilde P_a$. The optimal Bayesian strategy explains the data,  accounting for the larger boosting at small prevaccination titers and showing no increase for large titers, while the proportional model predicts constant boosting for all titers.
\AM{
Similar experimental results have been reported for antibody titers pre- and post a shingles vaccination \cite{Li2017}.
Mechanistic models have been proposed to explain how the population dynamics of expanding lymphocytes might give rise to non-proportional boosting for both B cells and T cells \cite{Bocharov2011,DeBoer2013,Zarnitsyna2016,Mayer2018}. 
}

\AM{
Interestingly, for T cells Quiel et al. \cite{Quiel2011} have shown that fold expansion to peak cell numbers in an adoptive transfer experiment depends on the initial number of T cells as a power law with exponent $\sim -1/2$.
That scaling, which is for the peak expansion, predicts more expansion at high precursor number than Eq.~\ref{eqfoldchange}, which is for memory production. This implies a nonlinear relationship between peak T cell level and memory production, which further suppresses memory production at high precursor numbers. This prediction might be could checked in experiments measuring memory production after infection clearance, as well as the expansion peak.
}

\ifthenelse{\boolean{pnas}}{\subsection*{Long-term dynamics of a well-adapting repertoire}}{\subsection{Long-term dynamics of a well-adapting repertoire}}

\begin{figure*}
    \centering
    \includegraphics[scale=0.9]{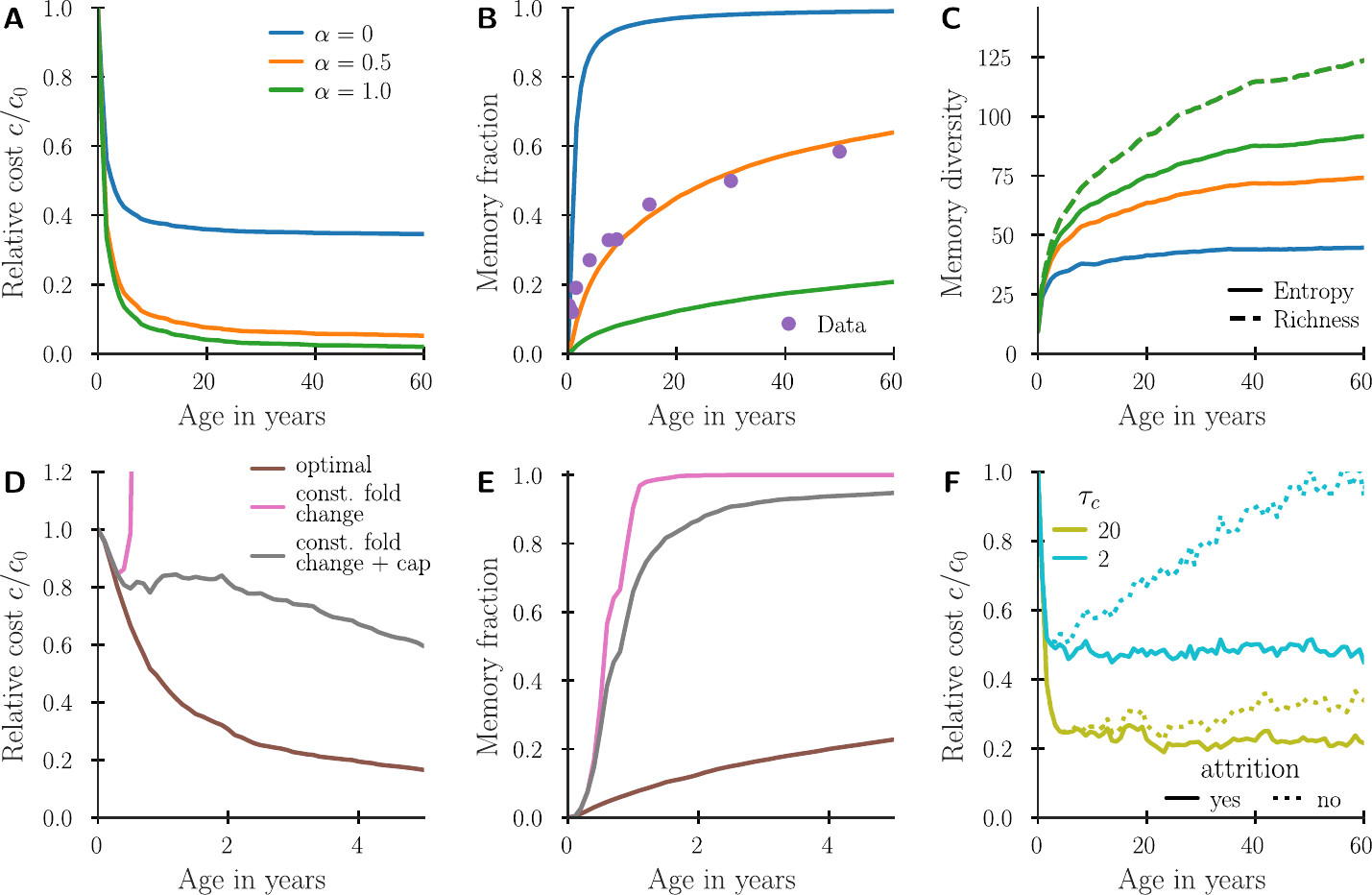}
    \caption{(A,D) Relative cost, (B,E) memory cell fraction, and (C) memory diversity as a function of age.
    (A-C) shows long term dynamics for a repertoire following optimal Bayesian update dynamics for three different $\alpha$.  Memory diversity is plotted as richness, i.e. the number of unique memory specificities, as well as the exponential of Shannon entropy $S$ of the memory compartement defined for a probability distribution $p_i$ with $\sum_i p_i = 1$ as $S = -\sum p_i \log p_i$. (D-E) compares the optimal Bayesian dynamics with a constant fold change update by a factor 30, and the same proportional update but with a cap to $10^5$ of the total fold expansion for any clone. (F) Comparison of the cost as a function of age for the optimal dynamics (solid lines) and a dynamics without attrition (dotted lines) for two environmental correlation time scales $\tau_c = 2 \tau/K \theta$ (see \siref~\ref{appspectralexpansion}).
    Parameters: encounter rate $\lambda = 40/\mathrm{year}$, antigen space dimensionality $K = 10^5$, antigen sparsity $\theta = 2.5 \cdot 10^{-4}$. In panel (A-E) we used $\tau \to \infty$, and in panel (D-F) $\alpha=0.5$. To reduce fluctuations the statistics were averaged over multiple runs of the dynamics.
    \AM{Data in (B) from \cite{Saule2006} and \cite{Shearer2003}.}
    \label{figagedependence}
    }
\end{figure*}

Our model makes predictions for the dynamics of growth and attrition of memory over time, with consequences for immunity and for the diversity of the immune repertoire.  We quantify the dynamics in terms of a {\it memory fraction} defined as a sum of the coverage fractions $\tilde P_{a_i}$ over all previously encountered pathogens $\{a_i\}$.    The memory fraction measures the size of memory relative to size of the whole immune repertoire.    Early in life every infection is new and even modest increases in the memory fraction lead to large drops in infection susceptibility (measured by the expected cost of new infections in Fig.~\ref{figagedependence}A).   At the same time, the memory fraction increases rapidly (Fig.~\ref{figagedependence}B), but the growth of memory slows as subsequent infections lead to less memory production following the optimal fold-change rule in Eq.~\ref{eqfoldchange}, and as attrition begins to play a role.   The fraction of the repertoire devoted to memory in mid-life is largely determined by how the cost of infections scales with coverage. The observed memory fraction of $\sim50\%$ at mid-life suggests  a cost exponent of  $\alpha \approx 0.5$ (Fig.~\ref{figagedependence}B).   The diversity of the memory repertoire increases with time at a rate that slows with age (quantified in Fig.~\ref{figagedependence}C by richness, which measures the number of unique specificities, and the Shannon entropy of the repertoire frequency distribution.)

To gain insight into these dynamics of our model we average the stochastic equations over the statistics of pathogen encounters.  
We show in \siref~\ref{appmeanfield} that this mean-field approximation yields a differential equation for the population fraction of different clones  with two opposing contributions which balance alignment of the immune repertoire with the current pathogenic environment (i.e. memory production) against alignment with the long-term mean environment (i.e. attrition). Interestingly, the mean-field equation broadly coincides with dynamics that were proposed in \cite{Mayer2015} to self-organize an optimal immune repertoire.  The essential difference here is that the time-scale of learning slows down with increasing experience following the rules of optimal sequential update in Eq.~\ref{eqfoldchange}.

We then asked which features of the proposed repertoire dynamics are most relevant to ensure its effectiveness. 
How important is the negative correlation between fold expansion and prior immune levels, and how important is attrition?
Furthermore,  if the immune system  follows Bayesian dynamics it must have integrated on an evolutionary time scale a prior about composition and evolution  of the pathogen environment through the parameters $\B \theta$ and $\tau$ -- however, the prior may be inaccurate.
How robust is the benefit of memory to imperfections of the host's prior assumptions about pathogen evolution?
To answer these questions we compare the long term immune repertoire dynamics using the optimal Bayesian scheme to other simplified schemes. 
We find that a constant fold expansion dynamics quickly leads to very suboptimal repertoire compositions (Fig.~\ref{figagedependence}D, pink line), since the exponential amplification of cells specific to recurrent threats quickly leads to a very large fraction of the repertoire consisting of memory of those pathogens (Fig.~\ref{figagedependence}E, pink line).   This  sub-optimality persists even if we assume that some global regulation  caps the constant fold expansion such that no individual receptor clone can take over all of the repertoire (Fig.~\ref{figagedependence}D,E grey line).  Thus,  negative feedback in T cell expansion to individual antigens is very important to maintain a properly balanced diverse repertoire. In contrast, within a dynamics with a negative correlation, the precise levels of updating do not need to be finely tuned to the environmental statistics: varying the assumed sparsity of the pathogen distribution, which controls fold expansion upon primary infection in the optimal dynamics, leads to a relatively modest deterioration of the convergence speed of the learning process (Fig.~\ref{figwrongprior}A) and does not matter asymptotically (Fig.~\ref{figwrongprior}B).
Attrition does not matter at young age, but can play an important role for long-term adaptation to relatively rapidly changing pathogen distributions (Fig.~\ref{figagedependence}F).  However,  the attrition time scale need not be finely tuned to get close to optimal dynamics  (Fig.~\ref{figwrongtimescale}).  
   
\ifthenelse{\boolean{pnas}}{\subsection*{Adapting a cross-reactive repertoire}}{\subsection{Adapting a cross-reactive repertoire}}

Above, we described adaptation of immune repertoires in terms of changes in  the effective coverage $\tilde P_a=\sum_r f_{a,r}P_r$, where the cross-reactivity matrix $F=(f_{a,r})$ reflects the ability of each receptor to recognize many antigens, and also the propensity of each antigen to bind to many receptors \cite{Mason1998}.   Because of cross-reactivity, each pathogen encounter should result in the expansion of not just one but potentially many receptor clones. Here we ask how the optimal immune response is distributed among clones with different affinities.

Following Perelson and Oster \cite{Perelson1979}, we will represent the interaction of receptors and antigens by embedding both in a high-dimensional metric recognition ``shape space'', where receptors are points surrounded by recognition balls. Antigens that fall within a ball's radius will be recognized by the corresponding receptor.  In this presentation $a$ and $r$ are the coordinates of antigens and receptors respectively and their recognition propensity depends on their distance, $f_{a,r} = f(|a - r|)$.

Earlier sections have already discussed the optimal dynamics of the coverage $\tilde P_a$, which is a convolution of the cross-reactivity matrix with the receptor clone distribution $P_r$.  Thus, the optimal dynamics of the clone distribution can be derived by deconvolving the cross-reactivity subject to the constraint that $P_r$ cannot be negative.  Carrying out this analysis in \siref~\ref{appcrossreactivity}, reveals a general qualitative phenomenon -- competitive exclusion between clones expressed in the repertoire and their close neighbors within the cross-reactivity radius (Fig.~\ref{figcrossreactivityDelta}, blue line).  This exclusion is not an assumption of the model, but rather stems from the optimal Bayesian theory. Given a receptor clone that covers one region of antigenic shape space, the global likelihood of detecting infections increases by placing other clones to cover other regions.    This can be shown analytically when cross-reactivity is limited, memory updates are small in magnitude, and the pathogen distribution is assumed to be uncorrelated (\siref~\ref{appcrossreactivity}).

In general, the frequencies of pathogens might be correlated in antigenic space, for example because mutations from a dominant strain give rise to  new neighboring strains. An optimally adapting immune system should incorporate such correlations as a prior probability favoring smoothness of the pathogen distribution.   Such priors work their way through the optimal belief update scheme that we have described, and weaken the competitive exclusion between  clones with overlapping cross-reactivity (Fig.~\ref{figcrossreactivityDelta}, orange line).

In general, when cross-reactivity is wide or the required clone fraction update is large, numerical analysis shows that achieving optimally predictive immunity after a pathogen encounter requires a global reorganization of the entire repertoire (Fig.~\ref{figlocalupdating}, blue line).  There is no plausible mechanism for such a large scale reorganization since it would involve up- and down-regulation even of unspecific clones.   However, in \siref~\ref{appcrossreactivity} we show that the optimal update can be well-approximated by changes just to the populations of specific clones with pathogen binding propensities $f_{a,r}$ that exceed a threshold.   The optimizing dynamics with this constraint exhibits strong competitive exclusion, where only the highest affinity clones proliferate, while nearby clones with lower affinity are depleted from the repertoire (Fig.~\ref{figlocalupdating}, orange line). \AM{The local update rule provides protection that comes within 1 percent of the cost achievable by the best global update.} Thus,  reorganization of pathogen-specific receptor clone populations following an infection, as seen in vertebrates, can suffice to achieve near-optimal predictive adaptation of the immune repertoire.

 % discussion
\ifthenelse{\boolean{pnas}}{\section*{Discussion}}{\section{Discussion}}

The adaptive immune system has long been viewed as a system for learning the pathogenic environment \cite{Farmer1986}.   We  developed a mathematical framework in which this notion can be made precise. \AM{In particular, we derived a procedure for inferring the frequencies of pathogens undergoing an immigration-drift dynamics and showed how such inference might approximately be performed by a plausible population dynamics of lymphocyte clones.} We also argued that the antigenic environment must be effectively sparse to be learnable with a realistic rate of pathogen encounters. The optimal repertoire dynamics in sparse antigenic environments naturally produces a number of known properties of the adaptive immune system including a large memory production in naive individuals, a negative correlation of memory production with preexisting immune levels, and a sublinear scaling with age of the fraction of the repertoire taken up by memory of past infections. 

Our framework is easily extended to incorporate further aspects of pathogen evolution, e.g. mutational dynamics  in  antigenic space.   Such dynamics will lead to correlations in the pathogen distribution which we showed will influence the structure of the optimal conjugate repertoire. In particular, the optimal response should  spread around the currently dominant antigens to also provide protection against potential future mutations. Hypermutations in B cells may play a role in this diversification, in addition to their known function of generating receptors with increased affinity for antigens of current interest. 
 It would also be interesting to extend our framework to other immune defense mechanisms, including innate immunity, where the role of memory has received recent attention \cite{Netea2016}.

Although our study was motivated by the adaptive immune system, \AM{some of our main results extend to other statistical inference problems. We have extended earlier results on exactly computable solutions to the stochastic filtering problem for Wright-Fisher diffusion processes \cite{Chaleyat-Maurel2007} to derive an efficient approximate inference procedure.
This procedure might be of use in other contexts where changing distributions must be inferred from samples at different time points, e.g., in population genetics. }Additionally, we have derived the convergence rate for Bayesian inference of categorical distributions in high dimensions in the undersampled regime, showing that effectively sparse distributions can be inferred much more quickly. These results add to the growing literature on high-dimensional inference from few samples \cite{Johnstone2009,Advani2016}, which has arisen in the context of the big data revolution.

We propose that the adaptive immune system balances integration of new evidence against prior knowledge, while discounting previous observations to account for environmental change.
Similar frameworks have been developed for other biological systems.  In neuroscience, leaky integration of cues has been proposed as an adaptive mechanism to discount old observations in change-point detection tasks \cite{Glaze2015, Glaze2018}, and close-to-optimal accumulation and discounting of evidence has been reported in a behavioral study of rat decision-making in dynamic environments \cite{Piet2017}.    Inference from temporally sparse sampling has been considered in the framework of infotaxis, which is relevant for olfactory navigation \cite{Vergassola2007}.     In the context of immunity, related ideas about inference and prediction of pathogen dynamics have been used to predict flu strain and cancer neoantigen evolution in silico  \cite{Luksza2014,Luksza2017}.   Finally, ideas similar to those developed here could be used in ecology or microbiome studies to reconstruct evolutionary or ecological trajectories of population dynamics from incomplete sampling of data at a finite number of time points, e.g., from animal sightings or metagenomics.

 % methods
\ifthenelse{\boolean{pnas}}{}{\section{Methods}}
\subsection{Modeling pathogen dynamics by a immigration-drift process}
In our model we describe the stochastic dynamics of the pathogenic environment (Fig.~\ref{figsetup}A) by a Fokker-Planck equation for the conditional probability distribution $\rho(\B Q, t)$
\begin{equation}
     \frac{\partial \rho (\B Q, t)}{\partial t} = \, \mathcal{A} \rho (\B Q, t),
\end{equation}
where $\mathcal{A}$ is a differential operator acting on $\rho$ that controls the dynamics.
For concreteness, we consider a population that changes due to genetic drift and immigration from an external reservoir, which we describe by a Wright-Fisher diffusion equation \cite{Etheridge2012,Griffiths2010}
\begin{equation} \label{eqwfdiffusion}
    \begin{split}
        \tau \frac{\partial \rho (\B Q, t)}{\partial t} = &-\frac{1}{2} \sum_{a} \frac{\partial}{\partial Q_a} \[[(\theta_a - |\B \theta| Q_a) \rho (\B Q, t) \]]  \\
                  &+ \frac{1}{2} \sum_{a,b} \frac{\partial^2}{\partial Q_a \partial Q_b} \[[ Q_a (\delta_{a, b} - Q_b) \rho (\B Q, t) \]],
    \end{split}
\end{equation}
where $\tau$ sets the time scale of  dynamics, $\B \theta$ is a $K$-dimensional vector of immigration rates, and $\delta_{a, b}$ is the Kronecker delta, which is 1 if $a = b$ and 0 otherwise. Here and in the following we denote the norm of a vector $\B x$ by $|\B x| = \sum_i x_i$. 
To efficiently simulate trajectories according to this dynamics we sample the new distribution of frequencies directly from the transition density of the stochastic process as described in App.~\ref{secwfsim}.
This dynamics retains key features of real pathogen environments.
First, at a given point in time the environment contains many different pathogens with different frequencies determined by genetic drift and immigration.    Second, the dominant pathogens change over time, such as is the case for many viruses e.g. the flu or HIV.

\subsection{Minimizing the cost of infection}

To solve the optimization problem Eq.~\ref{eqoptP} analytically a set of necessary conditions for optimality, the so-called Karush-Kuhn-Tucker conditions, can be derived.
When all receptors are present at a non-zero frequency in the optimal repertoire $P_r^\star > 0$ these conditions imply \cite{Mayer2015}
\begin{equation}
    \left. \frac{\partial \sum_a Q_a c(\tilde P_a)}{\partial P_r}\right|_{\B P^\star} = - \lambda^\star,
\end{equation}
where $\lambda^\star$ is set by the condition $\sum_r P_r^\star = 1$.
If we further simplify the problem by assuming that there is no cross-reactivity between different pathogens and by considering power-law cost functions then this simplifies to the explicit solution
\begin{equation} \label{eqoptP}
    \tilde P_a^\star = \frac{1}{Z} Q_a^{1/(1+\alpha)},
\end{equation}
where $Z$ is a normalization constant.
Other cases are discussed in detail in \cite{Mayer2015} including how to solve the optimization problem numerically using a projected gradient algorithm in the general case.

\AM{
\subsection{Change in protection upon a pathogen encounter}.
\label{methodsupdating}

The inference dynamics induces via the mapping from $\B{\hat Q}$ to $\B P^\star$ (Eq.~\ref{eqoptP}) a dynamics of an optimally adapting immune repertoire. To get intuition we derive how the coverage changes in a simple setting in which Eq.~\ref{eqoptP} holds (further cases are considered in \siref~\ref{appinducedrepdyn}.

By combining Eqs.\ref{equpdaten},\ref{eqavQ} we obtain an update equation for the expected frequencies upon encounter of antigen a as
\begin{equation} \label{eqhatqupdate}
    \B{\hat Q}^+ =  \frac{\B n^- + \B e_a}{|\B n^+|} = \frac{|\B n^-| \B{\hat Q}^- + \B e_a}{|\B n^+|},
\end{equation}
where to simplify notations we use $\B{\hat Q}(t^+) = \B{\hat Q}^+$, and where $|\B n^+| = |\B n^-| + 1$.
Using Eq.\ref{eqoptP} it follows that coverages are updated as
\begin{align}
    \tilde P_b^+ 
                 &=  \frac{[((\tilde P_b^-)^{1+\alpha} |\B n^-| + \delta_{a,b}/(Z^-)^{1+\alpha}) / |\B n^+|]^{1/(1+\alpha)}}{Z^+/Z^-}
\end{align}
Defining $\kappa = 1/ (|\B n^+| (Z^-)^{1+\alpha})$ and neglecting the change in normalization which is of order $1/K$ relative to the update size, we obtain Eq.~\ref{eqfoldchange}.
To fit the data set we note that a proportional rescaling of $\tilde P_a$ by a factor $k$ can be subsumed within the model by redefining $\kappa \to \kappa k^{1+\alpha}$. Therefore the scaling of $\tilde P_a$ to an antibody titer can be subsumed within $\kappa$.
}

{\bf Acknowledgements.}  % acknowledgement
The work was supported by grant ERCStG n. 306312, Simons MMLS grant 400425, and NSF grant PHY-1734030.  Work on this project at the Aspen Center for Physics was supported by NSF grant PHY-1607611.

\bibstyle{pnas}

\clearpage
\onecolumngrid

\appendix

\renewcommand\appendixname{SI Text}

\setcounter{table}{0}
\renewcommand{\thetable}{S\arabic{table}}%
\setcounter{figure}{0}
\renewcommand{\thefigure}{S\arabic{figure}}%

 % appwfdiffusion
\section{The immigration-drift model of pathogen dynamics}
\label{appwfdiffusion}

\subsection{Steady-state distribution and the backward equation}
The steady-state distribution of the pathogen dynamics (Eq.~\ref{eqwfdiffusion}) is a Dirichlet distribution with the parameter vector $\B \theta$ \cite{Etheridge2012},
\begin{equation} \label{eqdirichlet}
    \rho_s(\B Q)  = \frac{1}{Z(\B \theta)} \prod_i Q_i^{\theta_i-1} \eqqcolon \mathcal{D}(\B Q, \B \theta).
\end{equation}
The normalizing constant is the multivariate Beta function,  defined in terms of Gamma functions as $Z(\B \theta) = \prod_i \Gamma(\theta_i)/\Gamma(|\B \theta|)$.

\begin{figure}[h]
    \centering
    \includegraphics{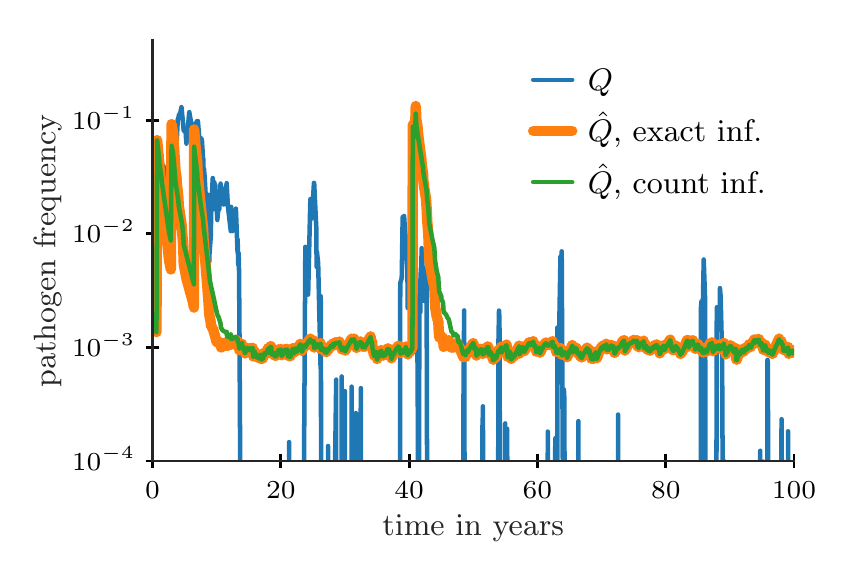}
    \caption{Frequency $Q$ of a pathogen over time in a dynamically changing environment along with inferred frequencies $\hat Q$. The approximate inference based on the count dynamics (Eqs.~\ref{equpdaten},\ref{eqpredn}) provides a very good approximation to the exact Bayesian inference implemented as described in App.~\ref{appspectralexpansion}.
    Parameter: $K=500$, $\theta=0.02$, $\tau = 10$ year, $\lambda = 10$/year.
    \label{figdyncomp}
    }
\end{figure}

To obtain a solution to Eq.~\ref{eqwfdiffusion} we 
write $\rho(\B Q, t)$ as the product of the steady-state distribution $\rho_s(\B Q)$ with a time-varying function $f(\B Q, t)$
\begin{equation} \label{eqrhoansatz}
    \rho(\B Q, t) = \rho_s(\B Q) f(\B Q, t).
\end{equation}
$f(\B Q, t)$ can be shown to obey the backward equation
\begin{equation} \label{eqbackward}
    \tau \frac{\partial f(\B Q, t)}{\partial t} = \frac{1}{2} \sum_{a} \[[(\theta_a - |\B \theta| Q_a) \frac{\partial}{\partial Q_a} f (\B Q, t) \]] \\
                  + \frac{1}{2} \sum_{a,b} \[[ Q_a (\delta_{a, b} - Q_b) \frac{\partial^2}{\partial Q_a \partial Q_b} f (\B Q, t) \]].
\end{equation}
By considering a coalescent dual process a probabilistic expansion can be derived from the backward equation (App.~\ref{appprobaexpansion}). This probabilistic expansion forms the basis of the approximate count dynamics Eqs.~\ref{equpdaten},\ref{eqpredn}. It also suggest an efficient method to sample from the transition density used in the simulation of the pathogen dynamics (Sec.~\ref{secwfsim}).
A decomposition of $f$ in Jacobi polynomials, which are the eigenfunctions of Eq.~\ref{eqbackward}, allows us to derive an efficient method to analytically solve the Bayesian prediction and update steps (App.~\ref{appspectralexpansion}). We use this alternative method to benchmark the quality of the inference achieved by the approximate counting scheme (Fig.~\ref{figdyncomp}).

 % appprobaexpansion
\subsection{Dirichlet mixture expansion of the belief distribution}
\label{appprobaexpansion}

Let us assume that the belief distribution at some time $B(\B Q, t)$ is a mixture of Dirichlet distributions
\begin{equation} \label{eqdirichletmixture}
    B(\B Q, t) = \sum_{\B m} c_{\B m}(t) \mathcal{D}(\B Q, \B m + \B \theta),
\end{equation}
with ${\B m} = (m_1,m_2,\cdots)$ for integer $m_i$ and mixture coefficients $c_{\B m} \geq 0, \sum_{\B m} c_{\B m} = 1$.
In particular this assumption holds true for the prior belief at $t=0$ which is described by the single component $\B m = 0$.
Both the update and prediction step of the inference procedure can be reduced to a much simpler procedure in terms of the mixture coefficients as derived below.
Using the mixture assumption $f(\B Q, t) = B(\B Q, t)/\rho_s(\B Q) $ is a mixture $f(\B Q, t) = \sum_{\B m} c_{\B m}(t) f_{\B m}(\B Q)$ of functions
\begin{equation} \label{eqfcounts}
    f_{\B m}(\B Q) = (|\B \theta|)_{|\B m|} \prod_i \frac{Q_i^{m_i}}{(\theta_i)_{m_i}},
\end{equation}
where $(a)_m = a(a+1)\cdots(a+m-1)$.
We can interpret the mixture coefficients as a probability distribution and define an average of a quantity $x$ as $\langle x \rangle = \sum_{\B m} c_{\B m} x_{\B m}$.
Using this notation the expected frequency is given by
\begin{equation}
    \bar{\B Q} = \langle (\B m + \B \theta)/(|\B m| + |\B \theta|) \rangle.
\end{equation}

To perform the update step Eq.~\ref{equpdate} we calculate
\begin{equation}
    Q_a f_{\B m}(\B Q) = \frac{\theta_a + m_a}{|\B \theta| + |\B m|} f_{\B m + \B e_a}(\B Q).
\end{equation}
Here and in the following we define $\B e_i$ as the unit basis vector with $i$-th entry 1 and all other entries 0.
It follows that the new belief distribution is still a mixture of Dirichlet distributions with
\begin{equation} \label{eqcmupdate}
    c_{\B m +\B e_a}(t^+) = \frac{\theta_a + m_a}{|\B \theta|+|\B m|} c_{\B m}(t^-) / Z,
\end{equation}
where $Z = \langle (\theta_a + m_a)/(|\B \theta|+|\B m|) \rangle_{c_{\B m}(t^-)}$ is a normalization constant.
To gain intuition about Eq.~\ref{eqcmupdate} let us consider the static environment limit $\tau \to \infty$. Applying Eq.~\ref{eqcmupdate} sequentially starting from the prior belief shows that the belief distribution remains composed of a single component ($c_{\B m} = 1$ for $\B m = \B m^\star(t)$, $c_{\B m} = 0$ otherwise) in this limit, where $\B m^\star(t)$ is the vector of counts of how often the different pathogens have been encountered. 
This is a classical result in Bayesian statistics \cite{Gelman2004}.
The fact that the belief distribution stays within the Dirichlet family when performing Bayesian updates starting from a Dirichlet prior is known as the conjugacy property of the Dirichlet prior with regards to the categorical likelihood function.
The interpretation of $\B m$ as the number of observations of a particular category motivates thinking about the $\theta_i$ as  pseudocounts, as according to Eq.~\ref{eqdirichletmixture} they are added to the counts of different categories to pull the estimate of the frequencies closer to the prior expectations.

The prediction step Eq.~\ref{eqpred} asks for an application of the backward operator $\mathcal{A}^\dagger$ to $f(\B Q, t)$.
As the backward equation is linear it acts independently on the different components.
Applying $\mathcal{A}^\dagger$ to $f_{\B m}$ yields
\begin{equation} \label{eqfmbackward}
    \mathcal{A}^\dagger f_{\B m}(\B Q) = \frac{1}{2} (|\B m | + | \B \theta| - 1) \[[\sum_{i=1}^K m_i f_{\B m-\B e_i}(\B Q) - |\B m| f_{\B m}(q)\]].
\end{equation}
By considering $\mathcal{A}^\dagger$ in Eq.~\ref{eqfmbackward} as acting on $\B m$ we obtain a dual death process description of the Wright-Fisher diffusion \cite{Griffiths2010}. Transitions from $\B m \to \B m - \B e_i$ happen at a rate $\frac{1}{2} (|\B m | + | \B \theta| - 1) m_i$.
The belief distribution continues to be a mixture of Dirichlet distributions with mixture coefficients that change as
\begin{equation}
    \tau \frac{\ud c_{\B m}}{\ud t} = \frac{1}{2} \(( |\B m | + | \B \theta|\)) \((\sum_{i} (m_i + 1) c_{\B m + \B e_i} \)) - \frac{1}{2} \(( |\B m | + |\B \theta| - 1\)) |\B m| c_{\B m}.
\end{equation}
Let us define the mean count vector $\langle \B m \rangle$. Its dynamical equation is
\begin{equation}
    \tau \frac{\ud \langle \B m \rangle}{\ud t} = \sum_{\B m} \B m \frac{\ud c_{\B m}}{\ud t}.
\end{equation}
Some algebra leads to the surprisingly simple yet exact equation
\begin{equation}\label{eqpredm}
    \tau \frac{\ud \langle \B m \rangle}{\ud t} = - \frac{1}{2} \langle ( |\B m | + | \B \theta| - 1) \B m \rangle.
\end{equation}

The probabilistic expansion allows the derivation of an efficient approximate scheme for inference. 
For peaked mixture distributions $c_{\B m}$ we can to a good approximation invert the order of calculating the expectation value and the product in Eq.~\ref{eqpredm}, i.e.
\begin{equation}
    \langle ( |\B m| + | \B \theta | - 1) \B m \rangle \approx ( |\langle \B m \rangle | + | \B \theta | - 1) \langle \B m \rangle.
\end{equation}
For peaked mixture distribution the update equation for the mean counts upon encountering pathogen $a$ is approximately
\begin{equation}
    \langle \B m(t^+) \rangle \approx \langle \B m(t^-) \rangle + \B e_{a},
\end{equation}
and the expected frequencies are approximately
\begin{equation}
    \B{\bar Q} \approx \langle \B m + \B \theta \rangle / (|\langle \B m + \B \theta \rangle|).
\end{equation}
Dropping the explicit notation for the average and replacing $\B n \coloneqq \B m + \B \theta$ we obtain Eqs.~\ref{equpdaten}--\ref{eqavQ} from the main text.

 % appspectralexpansion
\subsection{Spectral expansion of the belief distribution}
\label{appspectralexpansion}

As in App.~\ref{appprobaexpansion} we decompose the belief distribution but now we decompose $f(\B Q, t)$ into the eigenfunctions $g_{\B n}(\B Q)$ of the backward operator. This approach leads to simpler decoupled equations for the prediction step of the Bayesian inference.
To simplify notations we consider the case $K=2$ but the derivation generalizes to arbitrary $K$ \cite{Griffiths2010}.
Note that to describe the dynamics of the frequency of any particular pathogen, we can lump together the frequencies of the other $K-1$ pathogens for the model we consider. In particular, as the rates of immigration of different types are independent of composition of the population, the dynamics of the chosen pathogen does not depend on distribution of pathogens among the other types.
The dynamics of the $i$-th type for arbitrary $K$ can thus be mapped to $K=2$ by setting $\theta_1 \coloneqq \theta_i$ and $\theta_2 \coloneqq \sum_{j \neq i} \theta_j$.

The steady-state distribution specializes from the Dirichlet distribution to the Beta distribution for $K=2$
\begin{equation}
    \rho_s(q) = q^{\theta_1-1} (1-q)^{\theta_2-1}/Z,
\end{equation}
where $Z = B(\theta_1, \theta_2) = \Gamma(\theta_1) \Gamma(\theta_2)/\Gamma(\theta_1+\theta_2)$ is the Beta function.
The eigenfunction $g_{\B n}$ for an eigenvalue $\lambda_n$ needs to fulfill the backward equation (Eq.~\ref{eqbackward}), which leads to
\begin{equation} \label{eqbackwardeigen}
    \frac{1}{2} q(1-q) \frac{\ud^2 g_n}{\ud q^2}(q) + \frac{1}{2}(- \theta_2 q + \theta_1 (1-q)) \frac{\ud g_n}{\ud q}(q) = - \tau \lambda_n g_n(q),
\end{equation}
where $q=Q_1$.
Up to a rescaling this equation is a Jacobi differential equation.
The solutions of this differential equation are the modified Jacobi Polynomials 
\begin{equation}
    g_n(q) = P_n^{(\theta_2-1, \theta_1-1)}(2q-1),
\end{equation}
where $P_n^{(a, b)}(x)$ is the n-th Jacobi polynomial with eigenvalue
\begin{equation}
    \lambda_n = \frac{1}{2 \tau} n (n+\theta_1+\theta_2-1).
\end{equation}

Let us first state a number of properties of these polynomials which we will need later.
The polynomials form an orthogonal system with respect to the weight function $q^{\theta_1-1} (1-q)^{\theta_2-1}$, i.e.
\begin{equation} \label{eqorthogonality}
    \int_0^1 \ud q \, g_n(q) g_m(q) q^{\theta_1-1} (1-q)^{\theta_2-1} = \delta_{n,m} \Delta_n(\theta_1, \theta_2),
\end{equation}
where the normalization coefficients $\Delta_n(\theta_1, \theta_2)$ are given by
\begin{equation}
    \Delta_n(\theta_1, \theta_2) = \frac{\Gamma(n+\theta_1) \Gamma(n+\theta_2)}{(2n + \theta_1 + \theta_2 - 1) \Gamma(n + \theta_1 + \theta_2 -1) \Gamma(n+1)}.
\end{equation}
For $n \geq 1$ the polynomials are related by the recursion formula \cite{Song2012}
\begin{equation} \label{eqpolynomialrecursion}
    q g_n(q) = \phi^-_n g_{n-1}(q) + \phi^0_n g_n(q) + \phi^+_n g_{n+1}(q),
\end{equation}
with coefficients
\begin{align}
    \phi^-_n &= \frac{(n + \theta_1 - 1) (n + \theta_2 - 1)}{(2 n + \theta_1 + \theta_2 - 1) (2 n + \theta_1 + \theta_2 - 2)}, \\
    \phi^0_n &= \frac{1}{2} - \frac{\theta_2^2 - \theta_1^2 - 2 (\theta_2 - \theta_1)}{2 (2 n + \theta_1 + \theta_2) (2 n + \theta_1 + \theta_2 - 2)}, \\
    \phi^+_n &= \frac{(n + 1) (n + \theta_1 + \theta_2 - 1)}{(2 n + \theta_1 + \theta_2) (2 n + \theta_1 + \theta_2 - 1)},
\end{align}
while for $n = 0$
\begin{equation}
    q g_0(q) =  \phi^0_0 g_{0}(q) + \phi^+_0 g_1(q),
\end{equation}
with coefficients
\begin{align}
    \phi^0_0 &= \frac{\theta_1}{\theta_1+\theta_2}, \\
    \phi^+_0 &= \frac{1}{\theta_1+\theta_2}.
\end{align}

Now let us write $f(q, t)$ as a linear combination of the eigenfunctions $g_n$ of the backward equation as,
\begin{equation} \label{eqeigendecomposition}
    f(q, t) = 1 + \sum_{n=1}^\infty d_n(t) g_n(q),
\end{equation}
with time-varying coefficients $d_n(t)$.
The prediction step of the Bayesian inference leads through Eq.~\ref{eqbackwardeigen} to  a simple exponential decay of the $d_n$,
\begin{equation} \label{eqdprediction}
    \frac{\ud d_n(t)}{\ud t} = -\lambda_n.
\end{equation}
The update step of the Bayesian inference leads through Eq.~\ref{eqpolynomialrecursion} to
\begin{equation} \label{eqdrecursion}
    d_n(t^+) = \chi^-_n d_{n-1}(t^-) + \chi^0_n d_{n}(t^-) + \chi^+_n d_{n+1}(t^-),
\end{equation}
where $\chi^-_n,\chi^0_n,\chi^+_n$ are constants.
These constants are normalized versions of the coefficients $\phi$ that appear in Eq.~\ref{eqpolynomialrecursion},
\begin{equation}
    \chi^{x}_n = \begin{cases}
        \phi^{x}_n/(c^0_0 + d_{1}(t^-) c^-_1) & \text{if pathogen 1 encountered}, \\
        (\delta_{x,0}-\phi^{x}_n)/(1-c^0_0 - d_{1}(t^-) c^-_1) & \text{if pathogen 2 encountered},
    \end{cases}
\end{equation}
for $x$ in $-, 0, +$, and where $\delta_{x,0} = 1$ for $x=0$ and $\delta_{x,0} = 0$ otherwise.
For efficient numerical computation note that the update step can be performed as a matrix multiplication of the triadiagonal matrix which has $c^0_n$ along the diagonal, $c^-_{n+1}$ above the diagonal, and $c^+_{n-1}$ below the diagonal with the vector of coefficients $\B d(t^-)$ followed by a normalization step.
We finally note that the expected frequency is obtained from Eq.~\ref{eqeigendecomposition} as
\begin{equation}
    \langle q(t) \rangle = c_0^0 + c_1^- d_1(t).
\end{equation}

\AM{
The timescale over which the pathogen frequencies change is set by the slowest timescale of the stochastic dynamics, i.e.
\begin{equation}
    \tau_c = \frac{1}{\lambda_1} = \frac{2\tau}{\theta_1+\theta_2}.
\end{equation}
This result generalizes to the general case $K > 2$ as
\begin{equation} \label{deftauc}
    \tau_c = \frac{2\tau}{|\B \theta|}.
\end{equation}
}

 % appdynamicscaling
\subsection{Defining an effective timescale for learning in changing environments}
\label{appdynamicscaling}

The total number of counts $|\B n|$ follows a piecewise deterministic decay process interspersed by updates at random times. We approximate this stochastic dynamics by the deterministic equation,
\begin{equation}
    \frac{\ud |\B n|}{\ud t} = - \frac{1}{2\tau} ( | \B n | - 1) (|\B n | - | \B \theta|) + \lambda,
\end{equation}
where we have replaced the stochastic jumps in the counts do to pathogen encounters by a source-term $\lambda$ equal to the rate of such jumps.
This differential equation is separable and can be solved, which for $|\B n(0)| = |\B \theta|$ yields
\begin{equation}
    |\B n(t)| = \frac{|\B \theta|+1}{2} + \AM{\sqrt{\eta}} \tanh\(( \frac{t \sqrt{\eta}}{2\tau} + \artanh\((\frac{|\B \theta| - 1}{2 \sqrt{\eta}}\)) \)),
\end{equation}
for
\begin{equation}
    \eta = 2 \lambda \tau + \frac{(|\B \theta|-1)^2}{4}
\end{equation}
Considering the limit of long environmental correlation times, $2 \lambda \tau \gg \frac{(|\B \theta|-1)^2}{4}$, we obtain the simplified expression
\begin{equation}
    |\B n(t)| - |\B \theta| \approx \sqrt{2\lambda\tau} \tanh\((\frac{\lambda t}{\sqrt{2\lambda\tau}}\)).
\end{equation}
If we further take the limit $2 \lambda \tau \gg (\lambda t)^2$ of no attrition, then we recover the simple linear scaling of the expected number of encounters with time
\begin{equation}
|\B n(t)| - |\B \theta| \approx \lambda t.
\end{equation}
The number of remembered encounters $|\B n(t)| - |\B \theta|$ relative to the effective number of present pathogens $|\B \theta|$ controls how much memory improves protection (Fig.~\ref{figdynamicscaling}).

\subsection{Efficient simulation by sampling from the transition density}
\label{secwfsim}

The transition density of going from $\B x$ to $\B y$ in time $t$ for the Wright-Fisher diffusion can be written as \cite{Griffiths2010}
\begin{equation}
    f(\B x, \B y, t) = \sum_{|\B l|=0}^\infty q_{|\B l|}^{|\B \theta|}(t) \sum_{\{\B l: |\B l| fixed\}} \mathcal{M}(\B l, \B x) \mathcal{D}(\B y, \B \theta + \B l),
\end{equation}
where $\mathcal{M}(\B l, \B x)$ is the multinomial distribution
\begin{equation}
    \mathcal{M}(\B l, \B x) = \binom{|\B l|}{\B l} x_1^{l_1} \cdots x_K^{l_K},
\end{equation}
and where $q_{|\B l|}^{|\B \theta|}(t)$ are the transition functions of a dual pure death process.
This description of the transition density can be derived based on similar arguments to those made in App.~\ref{appprobaexpansion} \cite{Griffiths2010}. The death process describes the loss of unmutated lineages going backward in time through coalescence and mutations.
For small times $q_{|\B l|}^{|\B \theta|}(t)$ is asymptotically normal with mean $\mu(t)$ and variance $\sigma^2(t)$ \cite{Griffiths1984,Jenkins2015} 
\begin{align}
    \mu(t) &= \frac{2\eta}{t}, \\
    \sigma^2(t) &= \begin{cases}
        \frac{2 \eta}{t} (\eta+\beta)^2 \(( 1+\frac{\eta}{\eta+\beta} - 2\eta\))\beta^{-2},  &\beta \neq 0, \\
        \frac{2}{3t},  & \beta = 0,
    \end{cases}\\
    \text{where } \beta &= \frac{1}{2} (|\B \theta| - 1) t, \\
    \text{and } \eta &= \begin{cases}
        \frac{e^\beta}{e^\beta - 1}, &\beta \neq 0, \\
        1, & \beta = 0.
    \end{cases}
\end{align}
To sample from the transition density function we thus proceed in three steps: Generate a normally distributed random number $|\B l|$ according to the asymptotic distribution of $q_{|\B l|}^{|\B \theta|}(t)$. Then draw $\B l$ from $\mathcal{M}(l, \B x)$. Finally, draw $\B y$ from $\mathcal{D}(\B \theta + \B l)$.

 % appinducedrepdyn
\section{Induced repertoire dynamics}
\label{appinducedrepdyn}

\subsection{Dependence of fold change upon a pathogen encounter on sparsity}.
\label{appupdating}

\AM{
To understand how memory production depends on environmental sparsity we specialize Eq.~\ref{eqfoldchange} to the case of a uniform prior distribution.
We then have} $|\B n| \approx K \theta$ and $Z^- \approx \sum_a 1/K^{1/(1+\alpha)} = K^{\alpha/(1+\alpha)}$, which for $K\theta \gg 1$ leads to $\kappa = 1/(K^{1+\alpha} \theta)$. 
The fold change upon an encounter of a pathogen starting from a naive repertoire $\tilde P_a^- = 1/K$ thus depends as follows on the sparsity of the environment,
\begin{equation}
    \tilde P_a^+ / \tilde P_a^- = [1 + 1/\theta]^{1/(1+\alpha)}.
\end{equation}

\subsection{Mean-field dynamics}
\label{appmeanfield}
Besides the large changes of the naive repertoire upon a primary infection there are situations in which the inferred distribution is changing in a more continuous manner, e.g. updating in the limit of many previous samples, or the prediction step. We thus now ask how small changes in the expected frequencies of pathogens $\B{\hat Q}$ change the coverage $\B{\tilde P}$.
We assume that there is no cross-reactivity $f_{r,a}=\delta_{r,a}$, and consider power-law cost functions, where we have optimal receptor frequency distribution $P^\star_r = Q_r^\beta/Z$ with $\beta = 1/(1+\alpha)$. 
As a preliminary we calculate the Jacobian
\begin{align}
    \frac{\partial P_r^\star}{\partial \hat Q_{r'}} &= \delta_{r,r'}\frac{1}{Z} \frac{\partial \hat Q_r^\beta}{\partial \hat Q_r}  - \frac{Q_r^\beta}{Z^2} \frac{\partial Z}{\partial \hat Q_{r'}},\\
    &= \beta P_r^\star \[[ \frac{\delta_{r,r'}}{\hat Q_r} - \frac{P_{r'}}{\hat Q_{r'}}\]],
\end{align}
We can then show that the dynamics in terms of $P^\star_r$ follows,
\begin{align}
    \frac{\ud P^\star_r}{\ud t} &= \sum_{r'} \frac{\partial P_r^\star}{\partial \hat Q_{r'}} \frac{\ud \hat Q_{r'}}{\ud t},\\
    &= P_r^\star \[[ \beta \frac{\ud \ln \hat Q_r}{\ud t} - \sum_{r'} P_{r'} \beta \frac{\ud \ln \hat Q_{r'}}{\ud t}\]],
\end{align}
which is of the form of a replicator equation
\begin{equation}
    \frac{\ud P^\star_r}{\ud t}  = P_r^\star \[[f_r - \bar f\]]
\end{equation}
with ``fitness" $f_r = \beta \frac{\ud \ln \hat Q_r}{\ud t}$ and mean fitness $\bar f = \sum_{r'} P_{r'} f_{r'}$.

Based on this general result we now analyze the dynamics of the repertoire due to the sequential Bayesian filtering.
Equivalently to Eq.~\ref{eqhatqupdate} the change of inferred distribution $\Delta \B{\hat Q} = \B{\hat Q}^+ - \B{\hat Q}$ upon encountering antigen $a$ is given by
\begin{equation} \label{eqDeltaQ}
    \Delta\B{\hat Q} = \frac{\B e_a - \B{\hat Q}}{|\B n|+1},
\end{equation}
where $\B e_a$ is the unit vector with $a$-th entry one and all other zero.
Asymptotically for large $|\B n|$ every update has a small effect only, and we might consider a mean-field description. In this description we replace $\B e_a$ by its expectation value $\B Q$ and define an average rate of change per unit time by multiplying the update size by the frequency $\lambda$ of pathogen encounters:
\begin{equation}
    \frac{\ud \B{\hat Q}}{\ud t} = \frac{\lambda}{|\B n|+1} (\B Q - \hat{\B Q}).
\end{equation}
Here $\B Q$ is the actual distribution of pathogens, and $\hat{\B Q}$ are the expected frequencies of pathogens based on the immune system's internal belief.
For the prediction step we have a dynamics for counts, which we can convert into a dynamics for the inferred distribution.
We have $\hat Q_r = n_r / |\B n|$ and a dynamics on counts given by Eq.~\ref{eqpredn}. From there we obtain
\begin{align}
    \frac{\ud \hat Q_r}{\ud t} &= \frac{1}{|\B n|} \frac{\ud n_r}{\ud t} - \frac{n_r}{|\B n|^2} \frac{\ud |\B n|}{\ud t}, \\
    &= - \frac{(|\B n|-1)|\B \theta|}{2 \tau |\B n|}(\hat Q_r - \hat Q_r^0),
\end{align}
where $\hat Q_r^0 = \theta_r / |\B \theta|$ is the prior guess for the distribution.
Taken together we have
\begin{equation}
    \frac{\ud \B{\hat Q}}{\ud t} = \gamma(t) (\B Q - \hat{\B Q}) - \delta(t) (\B{\hat Q} - \B{\hat Q}^0).
\end{equation}
with the (time-varying) coefficients
\begin{align}
    \gamma(t) = \frac{\lambda}{|\B n(t)|+1},\\
    \delta(t) = \frac{(|\B n(t)|-1)|\B \theta|}{2 \tau |\B n(t)|}.
\end{align}

The fitness in the replicator equation is then
\begin{equation}
    f_r = \beta \((\gamma(t) \frac{Q_r}{\hat Q_r} + \delta(t) \frac{\hat Q_r^0}{\hat Q_r}\)).
\end{equation}
The fixed point of the dynamics in a static environment $\delta(t) = 0$ is the optimal repertoire as expected from the asymptotic optimality of Bayesian inference. Replacing $\hat Q_r = (Z P_r^\star)^{1+\alpha}$ we then obtain a fitness
\begin{equation}
    f_r = \frac{\gamma(t)}{Z^{1+\alpha}} \frac{Q_r}{(P_r^\star)^{1+\alpha}}
\end{equation}
which except for the prefactor is equivalent to the population dynamics proposed previously in \cite{Mayer2015}.
That work did not consider the prefactor that leads to a slowing down of the dynamics with time to reflect a tradeoff between new evidence and past experience.
The prediction steps relaxes the inferred distribution towards the prior distribution with a speed that for large $|\B n|$ is proportional to $|\B \theta|/\tau$.

\subsection{Updating a cross-reactive repertoire}
\label{appcrossreactivity}

We now consider the repertoire dynamics in the presence of cross-reactivity.
In a first order Taylor expansion the change in the repertoire composition upon a pathogen encounter is given by
\begin{equation} \label{eqDeltaPdef}
    \Delta \B P^\star \approx \B J_{\B G}(\B{\hat Q}) \Delta \B{\hat Q},
\end{equation}
where
\begin{equation}
    (\B J_{\B G})_{r,r'}(\B{\hat Q}) = \frac{\partial G_r(\B{\hat Q})}{\partial Q_r'}
\end{equation}
is the Jacobian of the mapping function (Eq.~\ref{eqoptP}).

The mapping between pathogen frequencies and the optimal repertoire takes the form $\B G(\B Q) = \B F^{-1} \B {\tilde P^\star}(\B Q)$ (if achievable given the constraint that no receptor frequency can be negative).
$\B{\tilde P^\star}(\B Q)$ is a function that depends on the cost function.
The Jacobian can thus be calculated using the chain rule as
\begin{equation}
\B J_{\B G}(\B{\hat Q}) = \B F^{-1} \B J_{\B {\tilde P^\star}}(\B {\hat Q}).
\end{equation}
For the power-law cost function we have $\tilde P^\star_a = R \hat Q_a^\beta/Z$, where $R = \sum_a f_{r, a}$ is the row sum of $F$, which assume to be constant.
Analogously to the derivation of the Jacobian in the previous section we derive
\begin{align}
    \frac{\partial \tilde P_a^\star}{\partial \hat Q_{a'}} &= \beta\tilde  P_a^\star \[[ \frac{\delta_{a,a'}}{\hat Q_a} - \frac{\tilde P_{a'}}{R \hat Q_{a'}}\]],
\end{align}
from which with some algebra follows
\begin{equation} \label{eqDeltaPfinal}
    \Delta\B P^\star \approx \((\frac{R}{\tilde P_a^\star}\))^\alpha\frac{1}{(1+\alpha) (|\B n|+1) Z^{1+\alpha}} \(( R \B F^{-1} \B e_a - \B P^\star \)).
\end{equation}
Here, there is a departure from the dynamics of the number $N_r$ of lymphocytes with receptor $r$ proposed in \cite{Mayer2015},
\begin{equation}
    \Delta N_r = N_r \Delta t \[[ A(\tilde N_a) f_{r,a} - d\]],
\end{equation}
where proliferation is proportional to $f_{r, a}$, instead of $(F^{-1})_{r,a}$.

 % appscaling
\section{Inference of high-dimensional categorical distributions from few samples}
\label{appscaling}

\subsection{Mean cost versus time}

In this Appendix we will derive analytical expressions for the optimized
cost as a function of time
\beq\label{eq:cost}
c(t)=\sum_{a=1}^K Q_a(t)c(\tilde P^\star_a(t))
\eeq
with the following simplifying assumptions: absence of
cross-reactivity, $f_{a,r}=\delta_{a,r}$ and $\tilde
P^\star_a(t)=P^\star_a(t)$; no attrition, $\tau\to+\infty$; and a power-law cost
function $c(P)=P^{-\alpha}$. The prior on $\B Q$ is a homogeneous Dirichlet
distribution: 
\beq
P(\B Q)\propto \prod_{a=1}^K Q_a^{\theta-1}.
\eeq

This problem is equivalent to the
Bayesian inference of a distribution drawn from a Dirichlet meta-distribution. Asymptotic
convergence properties of Bayesian inference procedures are
well-established \cite{Gelman2004}, but the convergence speed of
Bayesian estimators of the distribution in the non-asymptotic regime
has been much less studied to our knowledge. Analysing the
behaviour of $c(t)$ is equivalent to
analysing the convergence of the estimated distribution to the true
one with increasing number of samples. Here we will establish the
relevant scaling for few samples.

We consider the biologically relevant regime of high
dimension but effective sparsity of the distribution, $K \theta \gg
1$, $\theta \ll 1$.
Our main insight is that for such sparse distribution Bayesian inference is effective when the number of samples is on the order of a few $K \theta$, instead of the potentially much larger $K$.

The prominent role sparsity plays in allowing for more efficient estimation is reminiscent of compressed sensing \cite{Zdeborova2015}. Non-asymptotic results about inference in high-dimensional settings have been explored recently in the context of machine learning \cite{Advani2016}. Both connections merit further exploration.

We define the expected cost as $\<c(t)\>$, where the average is taken
over both random choices of $\B Q$, and random realizations of the
pathogen encounters, $\B n$, which are distributed according to:
\beq
n_a(t)\sim \mathrm{Poisson}(\lambda t Q_a),
\eeq
where $\lambda$ is the encounter rate. The number of encounters
determine the average belief for $\B Q$,
\beq
\hat Q_a(t)=\frac{\theta+n_a}{K\theta + \sum_{a=1}^L n_a}\approx
\frac{\theta+n_a}{K\theta + \lambda t},
\eeq
which itself shapes the optimal response and thus the cost through Eq.~\ref{eq:cost}:
\beq
P^\star_a(t)=\frac{\hat Q_a(t)^{1/(1+\alpha)}}{Z},
\eeq
where $Z$ is a normalization constant.

Note that the cost can be expressed in terms of a divergence
between the best receptor distribution $\B P^\star$ given full knowledge of $\B Q$ and the actual
receptor distribution $\B P$. Defining $P_a^\star=Q_a^{1/(1+\alpha)}/\tilde Z$ and replacing into Eq.~\ref{eq:cost}, we obtain:
\beq
c(t)=\tilde Z^{1+\alpha} \sum_a (P^\star_a)^{1+\alpha}P_a^{-\alpha}=c_\infty \exp\left[\alpha D_{1+\alpha}(\B P^\star\Vert \B P)\right],
\eeq
where $c_\infty = \tilde Z^{1+\alpha}$ is the asymptotic cost for $P = P^\star$ and where $D_{\beta}(\B P\Vert \B Q):=(\beta-1)^{-1} \ln[\sum_a P_a^\beta
Q_a^{1-\beta}]$ is the R\'enyi divergence of order $\beta$, which reduces to the
standard Kullback-Leibler divergence for $\beta=1$, i.e. $\alpha=0$.

\subsection{Reducing the problem to a single pathogen}

By symmetry all terms in the sum of Eq.~\ref{eq:cost} are equal on average and we have
\beq
\<c(t)\> = K \int dq\, \rho(q) q\, \<c(p)\>_{n},
\eeq
where we have introduced the short hand notations $q \coloneqq Q_a$
and $p \coloneqq \tilde P_a$, and where the average is taken over
$n\coloneqq n_a$. The pathogen frequency $q$ is approximately Gamma-distributed:
\beq
\rho(q)\approx \frac{(K\theta)^\theta}{\Gamma(\theta)} q^{\theta-1}e^{-K\theta q}.
\eeq

In general, the expectation value depends through $p$ on all previous
encounters with any of the pathogens (i.e. all the other $n_{a'}$,
$a'\neq a$). In high dimensions we can approximate this dependence by neglecting the correlation of the normalization factor $Z$ with $\hat q$ and using an effective $Z$.
For the power law cost functions we then have
\begin{equation} \label{eqp}
p = \hat q^{\frac{1}{1+\alpha}}/Z \quad \text{with} \; Z \approx K \langle \hat q^{\frac{1}{1+\alpha}} \rangle.
\end{equation}
For logarithmic cost this simplifies to $p = \hat q$ and $Z = 1$.
From Eq.~\ref{eqp} it follows that
\begin{equation}
\<c(t)\> \approx K Z^\alpha \langle q \hat q^{-\frac{\alpha}{1+\alpha}} \rangle = K^{1+\alpha} \langle \hat q^{\frac{1}{1+\alpha}}\rangle^\alpha \langle q \hat q^{-\frac{\alpha}{1+\alpha}} \rangle.
\end{equation}
We have $\hat q = (\theta + n)/(K\theta + \lambda t)$, but as the equation is invariant to a linear rescaling of $\hat q$ we can replace $\hat q$ by simply $\theta + n$.

\subsection{Costs for perfect or no information}
\label{appscalingextreme}

Asymptotically the distribution is learned perfectly and we have $\hat q = q$. Plugging this into the expressions derived previously we obtain
\begin{equation} \label{eqcinftyapprox}
    \bar c_\infty\coloneqq \<c(\infty)\> \approx K^{1+\alpha} \langle q^{1/(1+\alpha)} \rangle^{1+\alpha}, \quad \bar c_\infty \approx - K \langle q \ln(q) \rangle,
\end{equation}
for the power-law and logarithmic cost function respectively.
Performing the integrals we obtain
\begin{equation}
    \bar c_\infty \approx (K\theta)^\alpha \Gamma(1/(1+\alpha))^{1+\alpha}, \quad \bar c_\infty \approx \ln(K \theta) +\gamma,
\end{equation}
where $\Gamma(z)$ is the Gamma function and $\gamma$ the Euler-Mascheroni constant.
For $\alpha = 1$ this specializes to $\bar c_\infty = \pi K \theta$.
For the logarithmic cost $\bar c_\infty$ is equal to the Shannon entropy of the distribution, which suggests an interpretation of $K\theta$ as the effective number of pathogens that are present.

We can compare these costs for those obtained for a uniform repertoire 
\begin{equation}
    \bar c_0\coloneqq\<c(0)\> = K^\alpha, \quad \bar c_0 = \ln(K),
\end{equation}
to obtain
\begin{equation}
    \bar c_\infty/\bar c_0 \propto \theta^\alpha, \quad \bar c_\infty/\bar c_0 = (\ln(K\theta) + \gamma)/\ln(K)
\end{equation}
for power law and logarithmic cost respectively. As expected, in more sparse environments a larger relative improvement can be obtained by learning the distribution.

\subsection{Scaling in the  limit of few samples}
In the limit of small sampling, each pathogen has been seen at most once, meaning that $n$ is binary and distributed according to a Bernoulli variable with mean $\lambda t q$. Then we can use the approximation $\langle (\theta+n)^\beta \rangle \approx\theta^\beta (1-\lambda t q) + \lambda t q$ to obtain
\begin{align}
    \langle \hat q^{\frac{1}{1+\alpha}} \rangle &= \theta^{\frac{1}{1+\alpha}} + (1-\theta^{\frac{1}{1+\alpha}}) \lambda t \langle q \rangle \approx \theta^{\frac{1}{1+\alpha}} + \frac{\lambda t}{K} \\
    \langle q \hat q^{-\frac{\alpha}{1+\alpha}} \rangle &= \theta^{-\frac{\alpha}{1+\alpha}} \langle q \rangle + (1-\theta^{-\alpha}{1+\alpha}) \lambda t \langle q^2 \rangle \approx \frac{\theta^{-\frac{\alpha}{1+\alpha}}}{K} \left[ 1 - \frac{\lambda t}{K \theta}\right]
\end{align}
Putting things together we obtain
\begin{equation}
   \< c(t)\> \approx K^\alpha \left(1+ \frac{\lambda t}{K\theta}\theta^{\frac{\alpha}{1+\alpha}}\right)^\alpha \left( 1 - \frac{\lambda t}{K \theta}\right) \approx K^\alpha \left( 1 - \((1-\alpha \theta^{\frac{\alpha}{1+\alpha}}\))\frac{\lambda t}{K \theta}\right),
\end{equation}
which, except for a correction that vanishes as $\theta \to 0$, scales with $\lambda t/(K\theta)$.

For the logarithmic cost we approximate similarly $\langle \ln(\theta + n) \rangle \approx (1-\lambda t q) \ln(\theta) + \lambda t q \ln(1+\theta) \approx (1-\lambda t q) \ln(\theta)$.
We then have
\begin{equation}
    \<c(t)\> \approx \ln(K\theta + \lambda t) - K \ln(\theta) \[[ \langle q\rangle - \lambda t \langle q^2 \rangle \]].
\end{equation}
Using the formulas for the first and second moments we obtain
\begin{equation}
   \< c(t)\> \approx \ln(K\theta + \lambda t) - \ln(\theta) - \frac{\lambda t}{K\theta} \ln(1/\theta)
\end{equation}
Approximating further we have
\begin{equation}
    \<c(t)\> \approx \ln(K) \((1 - \((1-\frac{\ln(K\theta) - 1}{\ln(K)}\)) \frac{\lambda t}{K\theta} \)).
\end{equation}
Again the relative cost depends solely on $\lambda t /(K \theta)$ except for logarithmic corrections that vanish as $K \to \infty$ for fixed $K \theta$.

 % applogcost
\section{Infection cost in the expansion-delay regime}
\label{applogcost}

We have previously described mechanistic models that give rise to a power-law dependency of infection cost on the coverage \cite{Mayer2015}, in which we have assumed thatthe crucial determinant of infection cost is set by the time delay to recognition of the pathogen by the immune system. Experimental evidence shows that the initial recruitment of a large fraction of all specific lymphocytes often happens rapidly compared to the time it takes for the adaptive immune system to start clearing the infection \cite{Zehn2009}. We thus might hypothesize that the advantage of higher precursor numbers lies not in shortening the time to detection but in shortening the time to response by a sufficiently large number of effector cells. 

To derive the scaling of infection cost with coverage under these conditions, we consider that after an infection at time $0$, the number of specific cells grows exponentially with a rate $\gamma$, $N(t) = N(0) e^{-\gamma t}$. During the same time the pathogen population grows exponentially as well at a rate $\gamma_p$, $P(t) = P(0) e^{-\gamma_p t}$ until a time $t^\star$ at which a threshold level $N^\star$ of specific cells is reached. The expansion-delay time scales as $t^\star = \ln(N^\star/N(0))/\gamma$. If we assume that the cost of an infection is proportional $P(t^\star)$, then the cost scales as a power law with the initial number of specific cells $N(0)^{-\gamma_p/\gamma}$.

\clearpage
 % appfigures
\section{Supplementary figures}

\begin{figure*}[h]
    \begin{adjustbox}{center}
    \includegraphics{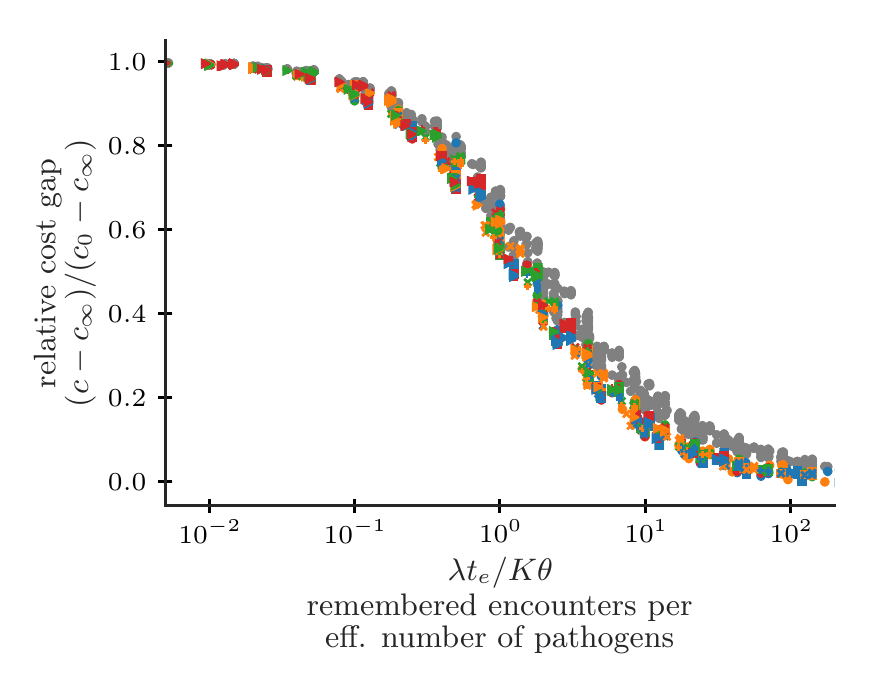}
    \end{adjustbox}
    \caption{Comparison of the dynamic scaling of the relative cost gap for $c(\tilde P_a) = \log \tilde P_a$ (colored) and $c(\tilde P_a) = 1/\tilde P_a$ (grey, see Fig.~\ref{figdynamicscaling}). For both cost functions rescaling parameters as $\lambda t_e / K \theta$ collapses the relative cost gaps for different parameters onto a similar master curve.
    \label{figdynamicscalingsi}
    }
\end{figure*}

\begin{figure*}[h]
    \begin{adjustbox}{center}
    \includegraphics{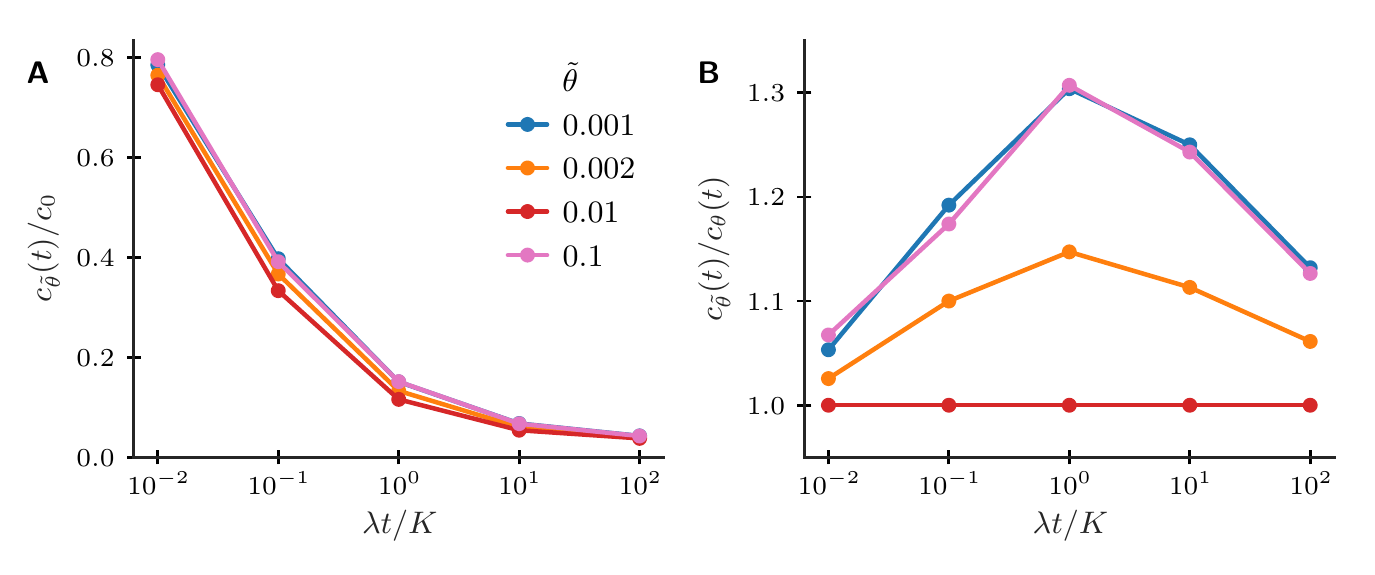}
    \end{adjustbox}
    \caption{Most of the benefit of memory is still achieved if the immune system has a slightly wrong prior about the antigen distribution sparsity $\theta$.
    (A) Scaling of mean cost of infections with different priors $\tilde \theta$ for a correct $\theta=0.01$ ($K=1000$).
    (B) Relative increase in cost by using a wrong prior vs age.
    \label{figwrongprior}
    }
\end{figure*}

\begin{figure*}[h]
    \begin{adjustbox}{center}
    \includegraphics{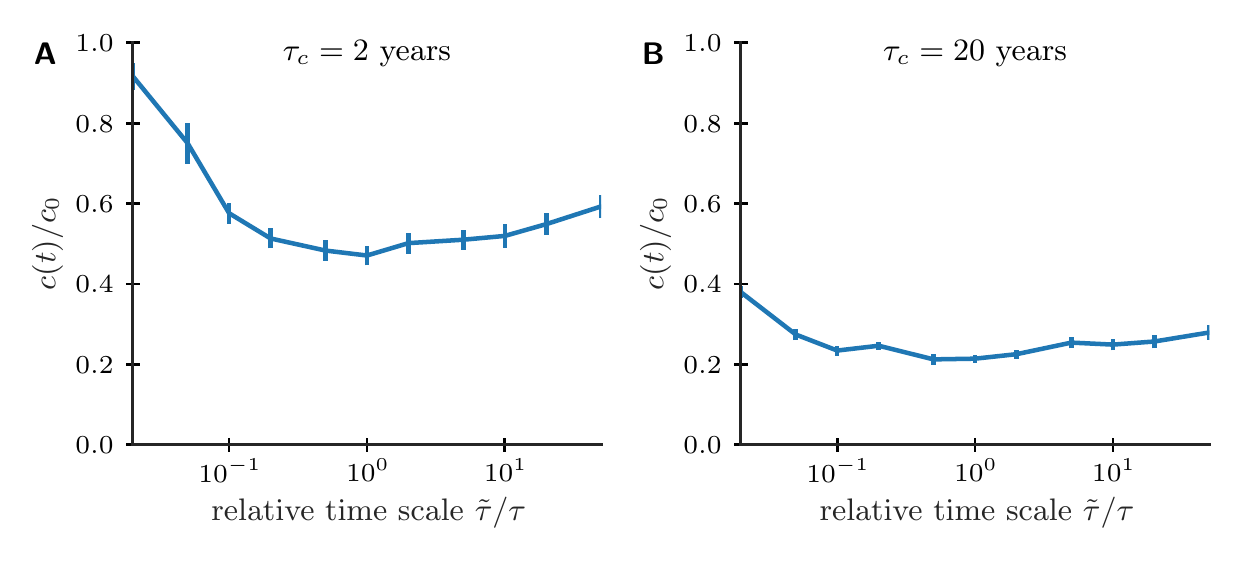}
    \end{adjustbox}
    \caption{The timescale of memory attrition does not need to be finely tuned for near-optimal prediction.
    Relative cost of infection at age 40 years as a function of the relative attrition timescale $\tilde \tau/\tau$ for environments with different correlation times $\tau_c = 2 \tau/K\theta$.
    Parameters: $\alpha=0.5$, $K$, $\theta$, $\lambda$ as in Fig.~\ref{figagedependence}.
    \label{figwrongtimescale}
    }
\end{figure*}

\begin{figure}[h]
    \begin{adjustbox}{center}
    \includegraphics{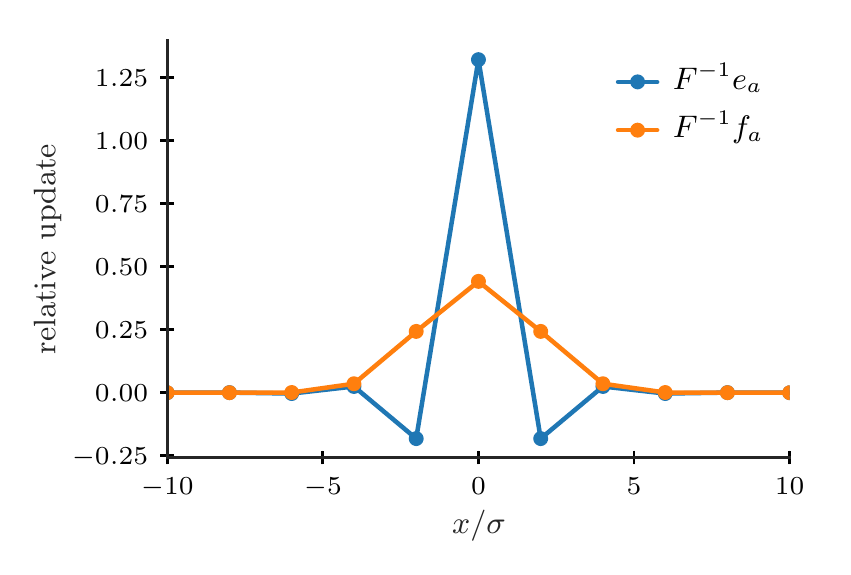}
    \end{adjustbox}
    \caption{While the optimal update of the receptor frequencies for uncorrelated pathogen environments (blue line) shows a depletion of close but suboptimal receptors such an effect might be reversed for correlated pathogen environments (orange line).  We use a Gaussian kernel to describe cross-reactivity $f(|a - r|) = e^{|a-r|^2/(2\sigma^2}$, and a spacing of receptors $0.5\sigma$. Assuming a change in inferred pathogen frequencies upon encountering pathogen $a$ of $e_a$ for the uncorrelated case (unit vector in $a$ direction) and $f_a = e^{(x-a)/(2\sigma_c^2)}$ with $\sigma_c=2 \sigma$ for the correlated case we obtain the optimal update in terms of receptor frequencies by multiplying the inverse of the cross-reactivity matrix $F$ with the change in frequencies. 
    \label{figcrossreactivityDelta}
}
\end{figure}

\begin{figure}[h]
    \begin{adjustbox}{center}
    \includegraphics{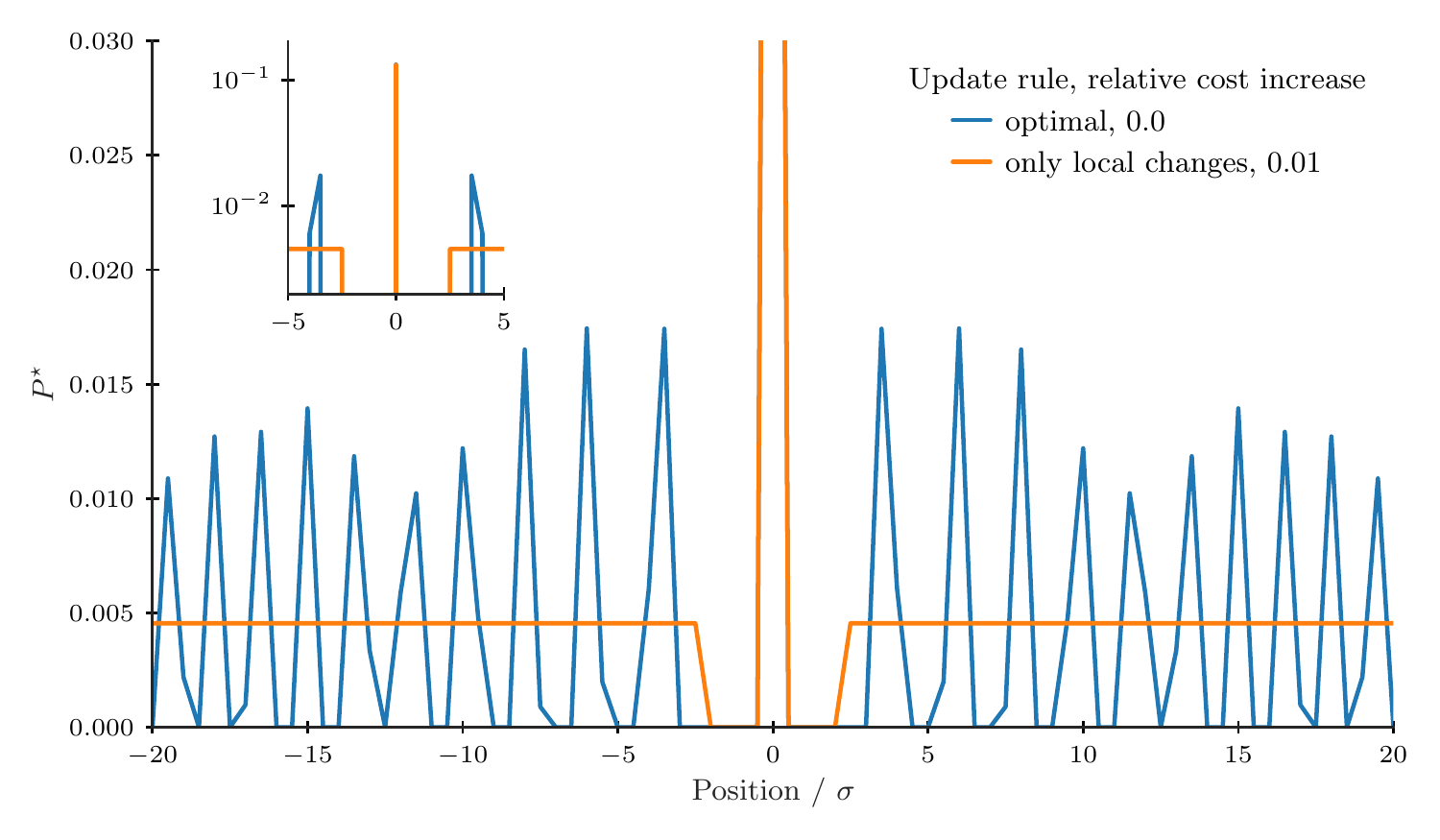}
    \end{adjustbox}
    \caption{For large changes in inferred frequencies the optimal repertoire can change drastically (blue line). A restriction to local changes yields a repertoire providing close to optimal protection, which exhibits a depletion of good, but suboptimal receptors (orange line).
            To obtain the global (local) optimal repertoire we use a projected gradient algorithm to minimize the cost function over all receptor distributions (over only the disribution of receptors within $2\sigma$ of the pathogen in shape space).
           We use a logarithmic cost function, $|\B \theta| = 10$, Gaussian kernel $f(|a - r|) = e^{|a-r|^2/(2\sigma^2)}$, and a spacing of receptors $0.5\sigma$.
    \label{figlocalupdating}
    }
\end{figure}

\end{document}